\documentclass[iop,apj,tighten,numberedappendix,twocolappendix]{emulateapj}

\usepackage[breaklinks,colorlinks,urlcolor=blue,citecolor=blue,linkcolor=blue]{hyperref}

\usepackage{amsmath,amssymb}
\usepackage{cleveref}
\usepackage{graphicx}
\usepackage{bm}
\usepackage[toc,title,page]{appendix}
\usepackage{tocbibind}
\usepackage{color}
\newcommand{\Msun}{{\rm M}_\odot}

\begin{document}
\title{Observable signature of merging stellar-mass black holes in 
active galactic nuclei
}

\author{
Hiromichi Tagawa\altaffilmark{1,2},
Shigeo S Kimura\altaffilmark{2,3},
Zolt\'an Haiman\altaffilmark{1},
Rosalba Perna\altaffilmark{4,5},
Imre Bartos\altaffilmark{6},
}

\affil{
\altaffilmark{1}Department of Astronomy, Columbia University, 550 W. 120th St., New York, NY, 10027, USA. \\
\altaffilmark{2}Astronomical Institute, Graduate School of Science, Tohoku University, Aoba, Sendai 980-8578, Japan. \\
\altaffilmark{3}Frontier Research Institute for Interdisciplinary Sciences, Tohoku University, Sendai 980-8578, Japan. \\
\altaffilmark{4}Department of Physics and Astronomy, Stony Brook University, Stony Brook, NY 11794-3800, USA. \\
\altaffilmark{5}Center for Computational Astrophysics, Flatiron Institute, New York, NY 10010, USA. \\
\altaffilmark{6}Department of Physics, University of Florida, PO Box 118440, Gainesville, FL 32611, USA.
}
\email{E-mail: ht2613@columbia.edu}

\begin{abstract} 
The origin of stellar-mass black hole mergers discovered through gravitational waves 
is being widely debated. 
Mergers in the disks of active galactic nuclei (AGN) represent a promising source of origin, with possible observational clues in the gravitational wave data.  
Beyond gravitational waves, a unique signature of AGN-assisted mergers is electromagnetic emission from the accreting black holes. Here we show that jets launched by accreting black holes merging in an AGN disk can be detected as peculiar transients by infrared, optical, and X-ray observatories. 
We further show that this emission mechanism can explain the possible associations between gravitational wave events and the optical transient ZTF19abanrhr 
and the proposed gamma-ray counterparts GW150914-GBM 
and LVT151012-GBM.  
We demonstrate how these associations, if genuine, can be used to reconstruct the properties of these events' environments. 
Searching for infrared and X-ray counterparts to similar electromagnetic transients in the future, once host galaxies are localized by optical observations, could provide a smoking gun signature of the mergers' AGN origin. 
 \end{abstract}

\section{Introduction}

Despite the large number of black hole mergers discovered by the LIGO \citep{2015CQGra..32g4001L}, Virgo \citep{2015CQGra..32b4001A} and KAGRA \citep{2021PTEP.2021eA101A} gravitational wave observatories \citep{Abbott21_GWTC3}, the astrophysical pathways to these mergers remain debated. A promising environment for mergers is the
disk of an AGN \citep{Bartos17,Stone17,McKernan19,Tagawa19}. 
Theory and observations \citep{Artymowicz1993,Levin2003,Tagawa19,Fan2022} 
suggest that black holes get embedded in AGN disks due to capture via dynamical interactions between the nuclear star cluster and the AGN disk \citep{Ostriker1983,Miralda-Escude2005}
and by in-situ star formation \citep{Levin2003,Milosavljevic2004}. 
The AGN disks then act as black hole assembly lines \citep{Cheng1999}, 
bringing the black holes closer together and helping them merge over relatively short time scales. It is also possible that multiple black holes merge consecutively, resulting in unusually heavy remnants \citep{Yang20_gap,Tagawa20_MassGap}. Comparisons to the observed black hole masses, spins and merger rate indicate that a sizable fraction of the observed mergers may originate in AGNs \citep{2021ApJ...920L..42G,Tagawa20_MassGap,Ford2022}. 
AGNs could also explain some of the peculiar detections, such as those with a high mass \citep{Tagawa20_MassGap} and possible high eccentricity \citep{Samsing20,Tagawa20_ecc,Romero-Shaw20,Gayathri2022,RomeroShaw2022}.

Due to the gas-rich merger environment, a key signature of the AGN channel is the possibility of electromagnetic emission accompanying the gravitational wave signal from the merger \citep{Bartos17,Stone17}. 
To explore this possibility, electromagnetic follow-up observations have been carried out for many of the mergers, with 
nine counterparts suggested so far. These include the peculiar optical flare ZTF19abanrhr with luminosity $L\sim10^{45}~{\rm{erg/s}}$ and duration $t_{\rm{duration}}\sim1~\rm{month}$, detected by the Zwicky Transient Facility (ZTF) in the AGN ${\rm{J124942.3+344929}}$. It was observed in spatial coincidence with, and within $18~\rm{days}$ of GW190521 \citep{Graham20,CalderonBustillo21}, 
the heaviest black hole merger detected to date \citep{LIGO20_GW190521}. 
Recently six optical flares possibly associated with massive black hole mergers with luminosity and duration similar to those of ZTF19abanrhr have been additionally reported by \citet{Graham2022}. 
The first gravitational wave event, GW150914 \citep{Abbott16a}, has also been proposed to have an associated bright gamma-ray event 
detected by the {\it Fermi} Gamma-ray Burst Monitor (GBM), GW150914-GBM \citep{Connaughton2016,Connaughton2018} with $L\sim10^{49}~{\rm{erg/s}}$ and $t_{\rm{duration}}\sim1~{\rm{s}}$ (but see \citealt{Greiner2016,Savchenko2016}). 
Finally, another gamma-ray burst candidate has been associated with the tentative gravitational wave event, LVT151012 \citep{Bagoly2016}, 
with properties similar to those of GW150914-GBM. 
As these associations remain debated and controversial \citep{Greiner2016,Savchenko2016,CalderonBustillo21}, 
it is necessary to consider whether a self-consistent physical model can, in fact, explain all of the properties of these tentative associations.

The possibility of electromagnetic emission from merging black holes in AGN disks has been actively studied recently, mainly focusing on optical flares \citep{Graham20,Kimura2021_BubblesBHMs,Wang2021_TZW,McKernan2019_EM}. 
Additionally, multiple scenarios have been proposed to produce emission similar to GW150914-GBM in more generic galactic environments \citep[e.g.][]{Janiuk2017,Perna2016}. 
However, no model has been able to explain all of the properties, including the luminosity, delay time, duration and color of electromagnetic transients. 
Furthermore, 
ZTF19abanrhr began brightening only approximately 18 days after the merger, which has not been physically 
justified (see \S~\ref{sec:ztf_model} below).

\begin{figure*}
\begin{center}
\includegraphics[width=160mm]{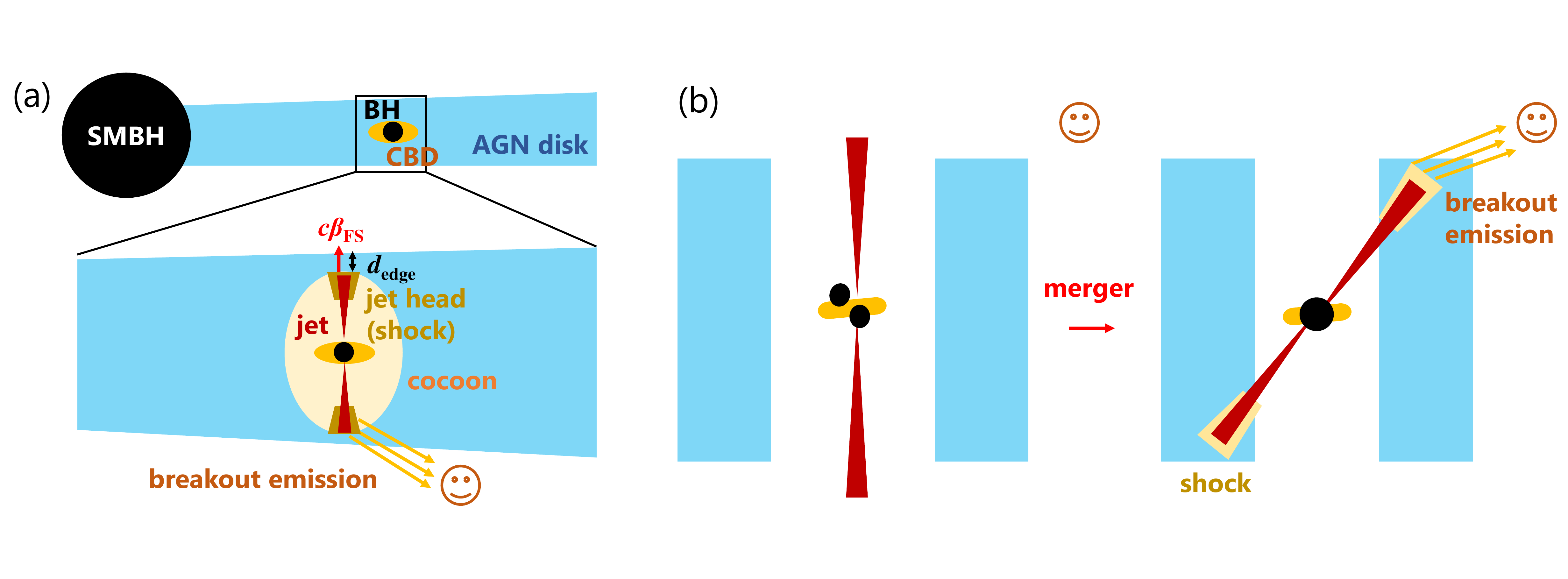}
\caption{
Schematic pictures of the breakout emission. 
(a) Breakout emission from the head of a jet launched from a solitary black hole embedded in an AGN disk. 
From a solitary black hole, the breakout emission is produced 
episodically, 
after replenishment of gas to the black hole \citep{Tagawa2022_BHFeedback}. 
(b) The emission from shocks produced around a jet after a merger. 
The jet direction is aligned with the black hole spin direction, which is reoriented by the merger. As a result, the jet again collides with unshocked AGN gas, producing breakout emission after the merger. 
}
\label{fig:schematic_solitary_merge}
\end{center}
\end{figure*}

A promising process for the production of detectable electromagnetic emission from AGN-assisted black hole mergers is the launch of a relativistic jet through the Blandford-Znajek effect (\citealt{Tagawa2022_BHFeedback}; see Fig.~\ref{fig:schematic_solitary_merge} for a schematic illustration). 
We evaluated the properties and observability of 
non-thermal and thermal emission from shocks emerging around 
such jets. 
In the AGN disk, a black hole is often surrounded by a circum-black hole disk \citep{Tagawa2022_BHFeedback}. 
When the circum-black hole disk is advection dominated, 
as realized in the slim disk model expected 
in this environment \citep{Abramowicz1988}, 
a magnetically dominated state can be realized owing to the accumulation of magnetic flux in the vicinity of the black hole (\citealt{Cao2011}; the magnetic field can be further magnified by an outflow; see e.g. \citealt{Liska2020}). 
In such cases, a jet is expected to be launched from a spinning black hole via the Blandford-Znajek process \citep{Blandford1977}. 

As the jet collides with unshocked gas in the AGN disk, strong forward and reverse shocks are formed in the disk and the jet head, respectively, and shocked material surrounds the sides of the jet (cocoon). 
During the early phases, photons in the shocked medium cannot escape from the system because they are surrounded by an optically thick disk. As the shock approaches the surface of the AGN disk, photons begin escaping from the system, leading to luminous thermal and non-thermal emission 
(Fig.~\ref{fig:schematic_solitary_merge}a). Note that solitary black holes produce similar breakout emission, but 
it will be difficult to observe because of 
lower brightness compared to that from merging black holes and 
the low duty-cycle of the breakout emission
 \citep{Tagawa2023_paper3} : the jet self-regulates to an episodic behaviour by creating a cavity around the black hole, with 
the breakout emission 
existing only for a small fraction of each cycle \citep{Kimura2021_BubblesBHMs,Tagawa2022_BHFeedback}. 
On the other hand, we show 
that breakout emission from merging black holes is 
expected to be observable due to its brightness, and that it accompanies up to several percent of black hole mergers in these environments (see $\S$\ref{sec:probability}). 
We further present a prescription for how non-thermal and thermal emission from these shocks can be used to "reverse engineer" the properties of the merger environments, applying the technique to the transients proposed to be associated with black hole mergers: ZTF19abanrhr, GW150914-GBM, and LVT151012-GBM.

\section{Model Description}
\label{sec:method}

In this section, we describe 
our model of accretion onto black holes, 
the properties of thermal and non-thermal emission, and our numerical choices. 
The various parameters of our model are summarized in 
Tables~\ref{table:parameter_fiducial} and \ref{table_notation} 
in Appendix~\ref{sec:notation}.
Readers not interested in the model details may skip directly to the next section, describing our results.

\subsection{Accretion onto black holes }
\label{sec:accretion_remnants}

A Blandford-Znajek jet is expected to be launched from rapidly accreting and spinning black holes in an AGN disk, 
as we outlined in Appendix~A.1 of \citet{Tagawa2022_BHFeedback}. 
The jet kinetic 
luminosity 
($L_{\rm j}$) is proportional to the mass accretion rate onto the black hole (${\dot m}$), 
\begin{align}
\label{eq:lj_macc}
L_{\rm j}=\eta_{\rm j}{\dot m} c^2,
\end{align}
where $\eta_{\rm j}$ is the jet conversion efficiency, which is approximated by $\eta_{\rm j}\sim a_{\rm BH}^2$ for a magnetically
dominated jet \citep{Tchekhovskoy2010,Narayan2021}, 
and $a_{\rm BH}$ is the dimensionless spin of the black hole. 
Since the power of a shock emerged around the jet and the luminosity of radiation emitted from the shock are roughly proportional to the jet 
kinetic 
luminosity, the accretion rate onto the black hole is a key quantity to determine the observed luminosity from the system.

The accretion rate onto 
a circum-black hole disk 
in the AGN disk is evaluated via the Bondi-Hoyle-Lyttleton rate, 
as given by Eq.~(1) of \citet{Tagawa2022_BHFeedback}. 
By considering the reduction from the Bondi-Hoyle-Lyttleton rate, 
we parameterized the fraction of the accretion rate onto the black hole (${\dot m}$) over the Bondi-Hoyle-Lyttleton rate (${\dot m}_{\rm BHL}$) as $f_{\rm acc}={\dot m}/{\dot m}_{\rm BHL}$. 
For example, low $f_{\rm acc}$ may be predicted due to winds from an accretion disk with a super-Eddington rate, although recent simulations suggest that the conversion to wind is moderate \citep{Kitaki2021} for accretion flows onto black holes in an AGN disk, in which the circularization radius is much larger than the trapping radius. 
In addition, the accretion rate onto the black hole in a cavity in active phases is estimated to be lower by a factor of a few compared to that without a cavity \citep{Tagawa2022_BHFeedback}. 
On the other hand, for merger remnants, 
inflow rates can be enhanced by shocks arising in the marginally bound annuli of the post-merger circum-black hole disks, due to recoil kicks imparted on the remnants. 
Considering this process, we adopt $f_{\rm acc}=15$ as a fiducial value, as discussed in detail in Appendix~\ref{sec:recoil_accretion}.

\begin{figure*}\begin{center}\includegraphics[width=160mm]{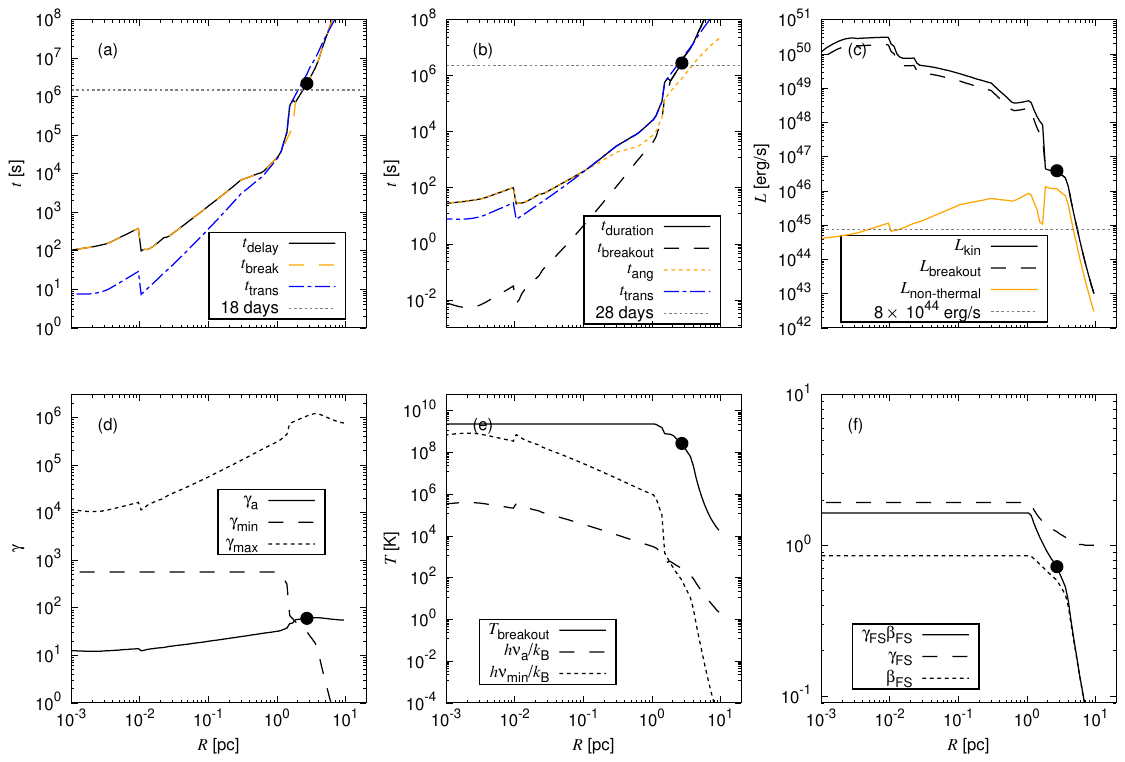}\caption{Various quantities as a function of the distance from the supermassive black hole ($R$) for breakout emission. (a)~The timescales representing delay ($t_{\rm delay}$, solid black), breakout ($t_{\rm break}$, dashed orange), and transparency ($t_{\rm trans}$, dotted blue). (b)~The observed duration of breakout emission ($t_{\rm duration}$, solid black), the duration of the shock breakout ($t_{\rm breakout}$, dashed black), the angular timescale ($t_{\rm ang}$, dotted orange), and the transparent timescale ($t_{\rm trans}$, dotted-dashed blue). 
(c)~The shock luminosity (solid black), the breakout luminosity by thermal emission (dashed black), and that by the non-thermal emission ($L_{\rm non-thermal}$, orange). 
(d)~The absorption ($\gamma_{\rm{a}}$, solid), minimum ($\gamma_{\rm{m}}$, dashed), and maximum ($\gamma_{\rm{max}}$, dotted) Lorentz factors. (e) The breakout temperature (solid), the absorption temperature ($h\nu_{\rm{a}}/k_{\rm{B}}$, dashed), and the minimum temperature ($h\nu_{\rm{m}}/k_{\rm{B}}$, dotted). (f) The shock velocity $\gamma_{\rm FS}\beta_{\rm FS}$ (solid), $\gamma_{\rm FS}$ (dashed), and $\beta_{\rm FS}$ (dotted). The black hole locations adopted in the fiducial model are indicated with filled circles superposed on the black solid lines. 
The dotted holizontal lines in panels~(a)--(c) present the observed values for $t_{\rm delay}$, $t_{\rm duration}$, and $L$ at $\sim2~{\rm{eV}}$. 
}\label{fig:prop_fid}\end{center}\end{figure*}

\subsection{Thermal emission at breakout }
\label{sec:breakout}

Here we provide details on the mechanisms involved in the production of thermal emission from shocks produced by the interaction between 
the AGN 
disk 
and the jet launched from the accreting black holes. 
We consider emission from the shocks propagating in optically thick media (an AGN disk, Fig.~\ref{fig:schematic_solitary_merge}). 
We assume that the black hole is at the midplane of the AGN disk, 
and the jet direction is perpendicular to the AGN disk plane. On the other hand, if the jet is inclined with respect to the angular momentum direction of the AGN disk by an angle $i$, the distance to the surface of the AGN disk from the black hole needs to be enhanced by a factor of $1/{\rm cos}(i)$ compared to the estimates in this paper, while other modifications would be minor. 
While the diffusion of photons is slower than the propagation of a shock ($v_{\rm FS}=\beta_{\rm FS}c>c/\tau$), where $\tau$ is the optical depth of the AGN disk, 
and $v_{\rm FS}=\beta_{\rm FS}c$ is the velocity of the forward shock, 
photons are trapped in the shock. 
Once the shock propagates to the height at which the diffusion to the edge (the disk photosphere) becomes faster than the propagation ($\beta_{\rm FS}c<c/\tau$), 
photons start to escape. 
The observed properties (such as the temperature, duration, and luminosity) of the emission from the shock breakout are 
differently characterized depending on the ranges of the shock velocity (Fig.~\ref{fig:prop_fid}), as described in the following \citep{Ito2020,Levinson2020}.

For Newtonian shocks with 
$\beta_{\rm FS}\gamma_{\rm FS}\lesssim0.03$ 
 \citep{Nakar2010,Sapir2013}, 
where $\gamma_{\rm FS}$ is the Lorentz factor of the forward shock, 
the radiation is in thermodynamic equilibrium 
as there is time to emit a sufficient number of photons by free-free emission. 
From the jump condition for a strong shock, 
the thermal equilibrium temperature is 
\begin{align}
\label{eq:tem_bo}
T_{\rm breakout}&\sim(18\rho_{\rm AGN}v_{\rm FS}^2/7a)^{1/4}\nonumber\\
&\sim10^4\,{\rm K}
\left(\frac{\beta_{\rm FS}}{0.02}\right)^{1/2}
\left(\frac{\rho_{\rm AGN}}{1\times 10^{-16}\,{\rm g~cm^{-3}}}\right)^{1/4}, 
\end{align}
where $a$ is the radiation constant 
and $\rho_{\rm AGN}$ is the density of the AGN disk. 
The breakout emission typically peaks in the optical or UV bands, $E_{\rm \gamma,BO}\sim2.8 ~k_{\rm{B}}T_{\rm breakout}\sim1$--$100$ eV, where $k_{\rm{B}}$ is the Boltzmann constant.

In non-relativistic regime, 
photons inside the shock start to diffuse out from the AGN disk, 
when the photon diffusion time, $t_{\rm diff}\sim~d_{\rm edge}^2\kappa_{\rm AGN}\rho_{\rm AGN}/c$, becomes equal to the shock expansion time, $t_{\rm dyn} \sim~d_{\rm edge}/v_{\rm FS}$,
where $d_{\rm edge}$ is the thickness of the AGN disk above the shock, and 
$\kappa_{\rm AGN}$ is the opacity of the AGN disk for thermal photons. 
By equating the timescales, the thickness at the breakout 
is given by 
$d_{\rm edge,BO}\sim~c/(v_{\rm FS}\kappa_{\rm AGN}\rho_{\rm AGN})$, and 
the duration of emission from a breakout shell is 
\begin{align}
\label{eq:t_bo}
t_{\rm breakout}
&\sim c/(v_{\rm FS}^2\kappa_{\rm AGN}\rho_{\rm AGN})
\nonumber\\
&\sim3\,{\rm yr}
\left(\frac{\beta_{\rm FS}}{0.1}\right)^{-2}
\left(\frac{\rho_{\rm AGN}}{1\times10^{-16}\,{\rm g~cm^{-3}}}\right)^{-1}, 
\end{align}
where we adopt the Thomson scattering opacity of $\kappa_{\rm AGN}\sim0.4\,{\rm g/cm^2}$ assuming ionized gas. 
The observed duration for bright emission ($t_{\rm duration}$) is typically given by $t_{\rm breakout}$ in non-relativistic cases.
The luminosity roughly evolves 
following $L\propto {\rm{e}}^{t^2}$ \citep{Sapir2011} and $L\propto t^{-4/3}$ 
before and after breakout during a planar phase ($t<t_{\rm sph}$), respectively, 
where $t_{\rm sph}=H_{\rm AGN}/v_{\rm FS}$ is the time of transition between the planar and spherical geometries of the breakout shell after the breakout of the shock 
and $H_{\rm AGN}$ is the scale height of the AGN disk. 
The dependence of the time evolution of the luminosity on the density profile of the ambient material is found to be weak \citep{Sapir2011}. 
After $t_{\rm sph}$, the luminosity evolves more modestly depending on the pre-shocked density profile. 
Note that for simplicity we adopt formulae obtained in a spherical geometry, while the evolution of the breakout luminosity from shocks propagating in a disk geometry would require modification from them \citep{Grishin2021}.

The delay time between the production of a jet and 
the breakout is roughly given by 
\begin{align}
\label{eq:t_break} 
t_{\rm break}&\sim \frac{3}{5} \frac{H_{\rm AGN}}{v_{\rm FS}}\nonumber\\
&\sim0.3\,{\rm yr}
\left(\frac{H_{\rm AGN}}{5\times10^{16}\,\mathrm{cm}}\right)
\left(\frac{\beta_{\rm FS}}{0.1}\right)^{-1},
\end{align}
where the prefactor of $3/5$ accounts for the deceleration during the non-relativistic regime. 
The time delay between the gravitational wave and the breakout emission ($t_{\rm delay}$) is roughly given by $t_{\rm break}$ in non-relativistic regimes. 
Note that $t_{\rm delay}$ can be lowered compared to $t_{\rm break}$ by up to the breakout timescale of gravitational waves ($t_{\rm GW,break}\sim H_{\rm AGN}/c$) if
the line-of-sight direction and the jet propagating direction coincide. 
This is because the breakout emission is produced closer to us in such cases compared to those for random jet directions.

The breakout luminosity of the thermal emission ($L_{\rm breakout}$) and 
the kinetic power of the shock ($L_{\rm sh}$) 
are typically \citep{Nakar2010} 
\begin{align}
\label{eq:l_bo}
L_{\rm breakout} &\sim 
L_{\rm sh}
\sim \pi\theta_{\rm j}^2H_{\rm AGN}^2\rho_{\rm AGN}
v_{\rm FS}^3\nonumber\\
&\sim 5\times 10^{43}\,{\rm erg/s}
\left(\frac{\theta_{\rm j}}{0.05}\right)^2  \nonumber\\
&\left(\frac{H_{\rm AGN}}{5\times 10^{16}\,\mathrm{cm}}\right)^{2}
\left(\frac{\rho_{\rm AGN}}{1\times 10^{-16}\,{\rm g~cm^{-3}}}\right) 
\left(\frac{\beta_{\rm FS}}{0.1}\right)^{3}, 
\end{align}
where $\theta_{\rm j}$ is the opening angle of the jet, 
which is given by $\theta_{\rm j}\sim (L_{\rm j} \theta_0^6 \beta_{\rm FS}^2 / \rho_{\rm AGN}H_{\rm AGN}^2 c^3)^{1/10}$ for non-relativistic regimes \citep{Bromberg2011}. 

For Newtonian shocks with 
$0.03 \lesssim \beta_{\rm FS}\gamma_{\rm FS}\lesssim1$, the breakout temperature ($T_{\rm breakout}$) strongly depends on the shock velocity \citep{Budnik2010,Sapir2013}. 
In these regimes, since the number of photons produced by free-free emission during propagation is lower than that required to establish thermal equilibrium \citep{Nakar2010}, 
the radiation is out of thermodynamic equilibrium and photons roughly follow local Compton equilibrium, whose radiation is characterized by a Wien spectrum $de_{\rm ph}/d\nu\propto\nu^3{\rm{e}}^{-h \nu/k_{\rm{B}}T_{\rm breakout}}$, 
where $\nu$ and $e_{\rm ph}$ are the frequency and 
the spectral energy density 
of photons, and $h$ is the Plank constant. 
The breakout temperature of the photons 
is calculated and fitted as \citep{Sapir2013} 
\begin{eqnarray}
\label{eq:tem_bo_fast}
{\rm log}_{10}
\left(\frac{T_{\rm breakout}}{{\rm 10^4~{\rm K}}}\right)
=0.975+1.735\left(\frac{\beta_{\rm FS}}{0.1}\right)^{1/2} \nonumber\\
+\left[0.26-0.08\left(\frac{\beta_{\rm FS}}{0.1}\right)^{1/2}\right]
{\rm log}_{10}\left(\frac{n_{\rm AGN}}{10^{15}~{\rm cm^{-3}}}\right), 
\end{eqnarray}
where $n_{\rm AGN}=\rho_{\rm AGN}/m_{\rm{p}}$ is the number density of the AGN disk gas 
and $m_{\rm{p}}$ is the proton mass. 
Note that this formula likely somewhat underestimates $T_{\rm breakout}$ for high $\beta_{\rm FS}$ and flat density profiles (roughly up to a factor of a few, Fig.~3 in \citet{Sapir2013}).

The duration of the breakout emission ($t_{\rm duration}$), 
the breakout timescale ($t_{\rm break}$), 
and the kinetic power of the shock ($L_{\rm sh}$) 
in the fast-Newtonian regimes are given by 
Eqs.~\eqref{eq:t_bo}, \eqref{eq:t_break}, and \eqref{eq:l_bo} as in the slow-Newtonian regimes.

We follow \citet{Nakar2012} to model evolution in relativistic regimes. 
For relativistic shocks with $\beta_{\rm FS}\gamma_{\rm FS}\gtrsim1$, 
electron/positron 
pairs are abundantly created, which enhances the production of photons inside the shock and regulates the temperature in the rest frame of the downstream plasma 
to $\sim100-200\,{\rm keV}$, almost independent of $\beta_{\rm FS}$ 
 \citep{Budnik2010}. 
Here, the relativistic shells are accelerated from the initial Lorentz factor of $\gamma_{\rm FS}$ to $\gamma_{\rm FS,f}=\gamma_{\rm FS}^{1+\sqrt{3}}$ due to the decrease of the gas density before breakout for $\gamma_{\rm FS}\lesssim4$ \citep{Johnson1971}, 
which is satisfied below. 
Note that breakout emission for $\gamma_{\rm FS}\gtrsim 4$ is not well investigated and applicability of the formalism is unclear. 
We also assume that the observer is within the angle of $\lesssim1/\gamma_{\rm sf,f}$ from the direction of the shock propagation, and that radiation is beamed, unless stated otherwise, 
where $\gamma_{\rm sf,f}\approx\gamma_{\rm sf}^{1+\sqrt3}$ is the final Lorentz factor of the shocked fluid, $\gamma_{\rm sf}$ is the initial Lorentz factor of the shocked fluid, 
and $\gamma_{\rm FS}\sim\sqrt{2}\gamma_{\rm sf}$ for strong shocks in relativistic regimes.

The delay time between a jet launch and 
the breakout from the AGN disk is roughly given by 
\begin{eqnarray}
\label{eq:t_break_rela}
t_{\rm delay}\sim 
{\rm{max}}\left(t_{\rm break,rel}, t_{\rm trans}\right), 
\end{eqnarray} 
where 
\begin{eqnarray}
t_{\rm break,rel}=\frac{H_{\rm AGN}}{4\gamma_{\rm FS}^2c} 
\end{eqnarray} 
is the breakout timescale of the shock in relativistic regimes, 
in which the factor of $1/2$ accounts for the deceleration during Blandford-McKee phases \citep{Blandford1976} 
and the factor of $H_{\rm AGN}/(2\gamma_{\rm FS}^2c)$ approximately accounts for the difference between the apparent travel time of the gravitational waves and the forward shock ($\sim~H_{\rm AGN}/(\beta_{\rm FS}c)-H_{\rm AGN}/c$). 
$t_{\rm trans}$ is the time at which the breakout shell 
with the temperature being $k_{\rm{B}}T_{\rm breakout}\sim200~{\rm keV}~\gamma_{\rm sf,f}$ in the shock immediate downstream 
becomes optically thin by the annihilation of pairs at the temperature of 
\begin{eqnarray}\label{eq:tem_bo_rela}
k_{\rm{B}}T_{\rm breakout}\sim50~{\rm keV}~\gamma_{\rm sf,f} \end{eqnarray} 
due to cooling of the shell by adiabatic expansion (we adopt Eq.~13 of \citet{Nakar2012}). 
The timescale $t_{\rm trans}$ should be taken into account for relativistic regimes since shocks are not transparent before the annihilation of pairs due to the enhancement of the optical depth by a factor of $\gtrsim100$.

The duration of the emission from the breakout shell is 
\begin{eqnarray}\label{eq:t_bo_rela}
t_{\rm duration}\sim {\rm{max}}
\left(t_{\rm ang}, t_{\rm trans},t_{\rm breakout}\right),
\end{eqnarray} 
where 
\begin{eqnarray}
\label{eq:t_ang}
t_{\rm ang}=\frac{H_{\rm AGN}}{2\gamma_{\rm sf,f}^2c}
\end{eqnarray} 
is the timescale during which radiation emitted by the same shell is observed due to its round shape
in the relativistic regime. 
In relativistic regimes and 
when shocks are propagating towards the observer, the breakout timescale ($t_{\rm breakout}$) is reduced due to the beaming effect 
by $\sim \gamma_{\rm FS}^2$, pair production by $100~\gamma_{\rm FS}$, and Klein-Nishina effect by $\gamma_{\rm FS}^{-2}$ \citep{Budnik2010}
as 
\begin{eqnarray}
\label{eq:t_bo_rel}
t_{\rm breakout}\sim \frac{c}{100v_{\rm FS}^2 \kappa_{\rm AGN} \rho_{\rm AGN}\gamma_{\rm FS}}. 
\end{eqnarray}

For relativistic cases, the kinetic power of the shock 
is \citep{Blandford1976}
\begin{eqnarray}
\label{eq:l_kin_rel}
L_{\rm sh} \sim 
\pi\theta_{\rm j}^2H_{\rm AGN}^2\rho_{\rm AGN}v_{\rm FS}\gamma_{\rm FS}^2c^2 ,
\end{eqnarray}
while the isotropic equivalent breakout luminosity 
is given by \citep{Nakar2012}
\begin{align}
\label{eq:l_bo_rel}
L_{\rm breakout} \sim 
\pi f_{\rm beaming}\theta_{\rm j}^2H_{\rm AGN}^2 
\rho_{\rm AGN}v_{\rm FS}\gamma_{\rm FS}^2c^2 \frac{\gamma_{\rm sf,f}}{4\gamma_{\rm sf}} 
\end{align}
where 
$f_{\rm beaming}$ is the beaming factor, 
takes into account the angular effect due to the fact that emission is concentrated in the direction of shock motion. 
We assume $f_{\rm beaming}=2\gamma_{\rm sf,f}^2$ approximating that the radiation is concentrated within the angle $\sim1/\gamma_{\rm sf,f}$ from the direction of the shock motion. 
Eq.~\eqref{eq:l_kin_rel} includes an additional boosting factor to account for conversion of the emitted power to the observer frame, to obtain the full Doppler boost. 
We assume that the breakout shell has a width of $v_{\rm FS}t_{\rm duration}$ (which is different from the width for the non-thermal emission of $v_{\rm FS}t_{\rm breakout}$). 
The factor of $\gamma_{\rm sf,f}/4\gamma_{\rm sf}$ in Eq.~\eqref{eq:l_bo_rel}
represents the reduction of the internal energy due to adiabatic expansion. 
The temperature in the immediate shock downstream is so high ($\sim200~{\rm keV}~\gamma_{\rm sf}$) that photons cannot escape from the shell because of copious electron/positron pairs. The photons escape from the shell after it expands and the temperature becomes $\sim50~{\rm keV}~\gamma_{\rm sf,f}$. Thus, the internal energy and the luminosity of the thermal emission is reduced by the factor of $\gamma_{\rm sf,f}/4\gamma_{\rm sf}$.

In summary, 
we adopt Eqs.~\eqref{eq:tem_bo}, \eqref{eq:tem_bo_fast}, \eqref{eq:tem_bo_rela} for $T_{\rm breakout}$ 
for $\beta_{\rm FS}\gamma_{\rm FS}\leq0.03$, 
$1 \geq \beta_{\rm FS}\gamma_{\rm FS}\geq0.03$, 
and $\beta_{\rm FS}\gamma_{\rm FS}\geq1$, 
respectively. 
The radiation is roughly characterized by 
a black-body radiation and a Wien spectrum 
for $\beta_{\rm FS}\gamma_{\rm FS}\leq0.03$ and 
$\beta_{\rm FS}\gamma_{\rm FS}\geq0.03$, 
respectively, 
although numerical simulations find that the spectrum at breakout is typically softer than that, especially for 
Newtonian slow shocks 
and in relativistic regimes \citep{Ito2020,Budnik2010}. 
Also, we assume that 
$t_{\rm delay}$ is given by 
Eqs.~\eqref{eq:t_break} and \eqref{eq:t_break_rela}, 
$t_{\rm duration}$ is given by 
Eqs.~\eqref{eq:t_bo} and \eqref{eq:t_bo_rela} for $\beta_{\rm FS}\gamma_{\rm FS}\leq1$ and $\beta_{\rm FS}\gamma_{\rm FS}\geq1$, respectively, 
and 
$L_{\rm sh}$
and $L_{\rm breakout}$ are given by 
Eqs.~\eqref{eq:l_bo} for $\beta_{\rm FS}\gamma_{\rm FS} \leq 1$, 
and are, respectively, given by 
Eqs.~\eqref{eq:l_kin_rel} and 
\eqref{eq:l_bo_rel} for $\beta_{\rm FS}\gamma_{\rm FS} \geq 1$.

The velocity of the shocked fluid ($\beta_{\rm sf}$) 
is determined by the competition between the power of the jet $L_{\rm j}$ and the mass of the ambient material swept up by the jet \citep{Bromberg2011}, which is derived using the density ($\rho_{\rm AGN}$), the scale height of the AGN disk ($H_{\rm AGN}$), and the injected opening angle of the jet ($\theta_{\rm j}$). 
Note that at the transitions where $\beta_{\rm sf} \gamma_{\rm sf}\sim1$ or ${\tilde L}\sim1$, where ${\tilde L}$ is the ratio between the energy density of the jet and the rest-mass energy density of the surrounding medium at the location of the head,
the formulae in \citet{Bromberg2011} unexpectedly predict that $\beta_{\rm sf}\gamma_{\rm sf}$ decreases as $L_{\rm j}$ increases. 
To avoid this, 
we set the upper limit for ${\tilde L}\leq1$ and the lower limit for ${\tilde L}\geq1$ to be $\beta_{\rm sf} \gamma_{\rm sf}=1$, 
since the two cases correspond to $\beta_{\rm sf}<1$ (with $\gamma_{\rm sf}\sim1$) and $\gamma_{\rm sf}>1$ (with $\beta_{\rm sf}\sim1$), respectively. 
Thus, $\beta_{\rm sf}\gamma_{\rm sf}$ is not accurately determined at around $\beta_{\rm sf}\gamma_{\rm sf}\sim1$.

To model ZTF19abanrhr, 
$\rho_{\rm AGN}$ and $H_{\rm AGN}$ are determined 
following the AGN disk model of \citet{Thompson05}, given the input parameters, the gas inflow rate (${\dot M}_{\rm in}$) and the angular momentum transfer parameter ($m_{\rm AM}$). Note that in the radii where the AGN disk is gravitationally unstable, $\rho_{\rm AGN}$ is proportional to $R^{-3}$, as the Toomre parameter $Q=\Omega^2/ (\sqrt{2} \pi \rho_{\rm AGN})$ becomes 1 (Eq.~3 in \citealt{Thompson05}), where $\Omega$ is the angular velocity of the merger remnant around the SMBH. 
$H_{\rm AGN}$ is determined to establish a stable state of the disk following the model of \citet{Thompson05}. 
On the other hand, 
to model GW150914-GBM and LVT191012-GBM, 
we determined $\rho_{\rm AGN}$ and $H_{\rm AGN}$ so that the observed properties of the flares are reproduced. 
Note that $\rho_{\rm AGN}$ and $H_{\rm AGN}$ are key parameters determining the observed properties, $t_{\rm delay}$, $t_{\rm breakout}$, and $L_{\rm breakout}$. 

To estimate the properties of breakout emission from the shock, 
we assume that the breakout velocity is given by the head velocity of the shocked region, 
although the directions of the jets may be random \citep{Tagawa20b_spin} and the side of the shock (cocoon) may first break out. 
Since the head velocity is $\sim5(\theta_{\rm c}/0.2)^{-1}$ times faster than the cocoon expansion velocity \citep{Bromberg2011}, where $\theta_{\rm c}$ is the opening angle of the cocoon, 
we assume that the cocoon can break out faster than the jet head when the angle between the propagation direction of the jet head and the AGN disk plane is less than $\theta_{\rm c}$. 
Assuming that the jet direction is random, 
the probability that the shock breaks out the AGN surface from its head is $\sim \int_{0}^{\pi/2-\theta_{\rm c}}{\rm sin}(\theta)d\theta/ \int_{0}^{\pi/2}{\rm sin}(\theta)d\theta\sim1-\theta_{\rm c}\sim0.8$ 
for $\theta_{\rm c}=0.2$. 
Thus, the shock emission most likely breaks out the AGN disk surface from the head, and we only consider this case.

\subsection{Non-thermal emission }
\label{sec:synchrotron}

In this subsection, we summarize the properties of the non-thermal emission emerged after the shock breakout. 
While photons are trapped in the shock ($\beta_{\rm FS}c>c/\tau$), 
the shock is mediated by scattering with the photons, 
in which electrons are not (or inefficiently) accelerated \citep{Ito2020}. 
Once photons start to escape, 
a collisionless shock can be formed, 
where electrons can be accelerated \citep[e.g.][]{Kashiyama2013b,Ito2020_Wind}.
These electrons produce non-thermal emissions via synchrotron and inverse Compton scattering processes.
Here we focus on non-thermal emission after the shock breakout timescale, which is given by $t_{\rm break}$ (Eq.~\ref{eq:t_break}) for non-relativistic and $t_{\rm break,rel}$ (Eq.~\ref{eq:t_break_rela}) for relativistic regimes, respectively. 
We focus on the non-thermal emission around the peak of the lightcurve.
The quantitative estimate of the time evolution would require numerical simulations, which is beyond the scope of this paper.

After the breakout of the shock, the non-thermal luminosity decreases depending on the density profile of the AGN disk and the speed of the shock. 
If the non-thermal radiation emerges -- and hence it is observed -- over a timescale longer than the intrinsic duration of particle acceleration, its brightness will be correspondingly diluted by a factor which we indicate with $f_{\rm dilution}$. 
For non-relativistic shocks ($\beta_{\rm FS}\gamma_{\rm FS}\leq1$), we have $f_{\rm dilution}= 1$ since the two timescales are identical. 
For relativistic shocks in gas-pressure dominated regimes for the AGN disk, since the gas density gradually changes as a function of the distance from the AGN disk plane, the luminosity gradually decreases on the observational timescale, and hence we again assume $f_{\rm dilution}=1$. 
On the other hand, for relativistic shocks in radiation-pressure dominated regimes for the AGN disk (which is realized for $R\lesssim 2\times 10^{-2}~{\rm pc}$ in the fiducial model, e.g. \citealt{Haiman2009}), the gas density abruptly decreases at a few $H_{\rm AGN}$ \citep[e.g.][]{Grishin2021}. 
In such cases, the non-thermal luminosity also abruptly decreases 
and is high only within the timescale of $\sim t_{\rm breakout}$.  In this case the  observed emission is diluted on the timescale  $t_{\rm dur,NT}={\rm max}(t_{\rm ang},t_{\rm breakout})$. 
By assuming for simplicity that the gas density above the AGN disk is almost in a vacuum, 
we then set $f_{\rm dilution}=t_{\rm breakout}/t_{\rm dur,NT}$ for the cases with relativistic shocks in radiation-pressure dominated regimes. 
We should note that the assumption may underestimate the non-thermal emission,  because disk winds might exist that increase the value of $f_{\rm dilution}$.

We assume that a fraction $\epsilon_{\rm{e}}$ of the kinetic energy of the shock is used to accelerate electrons in a collisionless shock 
 \citep{Sironi_2013}. 
Then, the synchrotron luminosity is given by \citep{Fan2008}
\begin{eqnarray}
\label{eq:l_sync}
L_{\rm syn} 
&\sim&
\frac{L_{\rm non-thermal}}{1+Y_{\rm SSC}+Y_{\rm 2ndIC}} \nonumber\\
&\sim &\epsilon_{\rm{e}}f_{\rm beaming}f_{\rm dilution}\frac{L_{\rm sh} }{1+Y_{\rm SSC}+Y_{\rm 2ndIC}}
\end{eqnarray}
where 
$Y_{\rm SSC}$ and $Y_{\rm 2ndIC}$ are the powers of synchrotron-self Compton and 
second order inverse Compton scattering compared to that of synchrotron emission, 
and $L_{\rm non-thermal}$ is the luminosity by non-thermal emission. 
We ignore the inverse Compton scattering of 
the thermal photons of the AGN disk and 
those of breakout emission. 
The former is because its energy density is much lower
compared to that of the synchrotron photons. 
The latter is because 
the scattering of such photons are typically in the Klein-Nishina regimes in our case. 
For $\gamma_{\rm sf}\beta_{\rm sf}\leq 1$, we set $f_{\rm beaming}=1$, and otherwise $f_{\rm beaming}=2\gamma_{\rm sf,f}^2$.

Here, plasma and/or MHD instabilities are assumed to amplify the magnetic field to $\epsilon_{\rm{B}}\lesssim10^{-3}$--$10^{-1}$, 
while electrons are accelerated via the first-order Fermi process with an energy fraction of 
$\epsilon_{\rm{e}}\lesssim10^{-2}$--$0.3$
(from observations, e.g., \citealt{Panaitescu2001,Frail2005},
and from theoretical studies, e.g., 
\citealt{Spitkovsky2008b,Sironi_2013}). 
We assume that electrons are accelerated in the shock to a power-law distribution of Lorentz factors $\gamma'_{\rm{e}}$ as 
$N(\gamma_{\rm{e}}')d\gamma_{\rm{e}}'\propto {\gamma_{\rm{e}}'}^{-p}d\gamma_{\rm{e}}'$ 
with a minimum ($\gamma_{\rm{m}}'$) and a maximum value ($\gamma_{\rm{max}}'$) and a power-law index $p$, 
where primes are used for quantities in the fluid comoving frame. 
Assuming that a fraction $\epsilon_{\rm{e}}$ of the kinetic energy 
is converted to electron energy, 
the minimum Lorentz factor $\gamma_{\rm{m}}'$ is given by 
\begin{align}
\label{eq:gamma_min}
\gamma_{\rm{m}}' \sim 
\epsilon_{\rm{e}}
\left(\frac{p-2}{p-1}\right) \frac{m_{\rm{p}}}{m_{\rm{e}}}(\gamma_{\rm sf}-1)
\sim 
40~\left(\frac{\epsilon_{\rm{e}}}{0.3}\right) 
\left(\frac{\gamma_{\rm sf}-1}{0.2}\right),
\end{align}
where $m_{\rm{e}}$ is the electron mass.

By comparing the cooling timescale by the synchrotron radiation and the acceleration timescale by the first-order Fermi acceleration mechanism, 
the maximum Lorentz factor of electrons is determined as 
\begin{align}\label{eq:gamma_max}
\gamma_{\rm{max}}'
&=\frac{v_{\rm sf}}{c}
\left(\frac{9\pi e}{10\sigma_{\rm T}B'_{\rm sf} \xi}\right)^{1/2}
\nonumber\\
&\sim 1\times 10^6 
\xi^{-1/2}
\left(\frac{\beta_{\rm sf}}{0.5}\right)
\left(\frac{A_{\gamma_{\rm sf}}}{1.5}\right)^{-1/4} \nonumber\\&
\left(\frac{\epsilon_{\rm{B}}}{0.1}\right)^{-1/4}
\left(\frac{\rho_{\rm AGN}}{1\times 10^{-16}\,{\rm g~cm^{-3}}}\right)^{-1/4}
\end{align}
where $e$ is the electric charge, $A_{\gamma_{\rm sf}}=(\gamma_{\rm sf}-1) (4\gamma_{\rm sf}+3)$,
\begin{align}
\label{eq:b_sf}
B'_{\rm sf}&=(8\pi\epsilon_{\rm{B}}e'_{\rm sf})^{1/2}\nonumber\\
&\sim 6 \times 10^2~{\rm G}~
\left(
\frac{A_{\gamma_{\rm sf}}
}{1.5}\right)^{1/2}
\left(\frac{\epsilon_{\rm{B}}}{0.1}\right)^{1/2} 
\left(\frac{\rho_{\rm AGN}}{1\times10^{-16}\,{\rm g~cm^{-3}}}\right)^{1/2},
\end{align}
is the magnetic field in the shocked medium 
(parameterized in terms of the
fraction $\epsilon_{\rm{B}}$ of postshock energy carried by the post shock magnetic field), 
$e'_{\rm sf}
=A_{\gamma_{\rm sf}}\rho_{\rm AGN}c^2$
is the internal energy density of the shocked medium, 
and 
$\xi$ is the parameter representing the ratio of the mean free path to the Larmor radius of electrons. 
We use $\xi=1$, but this choice only affects $\nu_{\rm{max}}$ and has little impact on most of the conclusions in this paper.

The cooling Lorentz factor of electrons, with which electrons can cool in the dynamical timescale due to radiation, is given by 
\begin{eqnarray}
\label{eq:gamma_c}
\gamma'_{\rm c}={\rm{max}}\left(1,\frac{6\pi m_{\rm{e}} c}{\sigma_{\rm T}B_{\rm sf}'^2\gamma_{\rm sf}t_{\rm break}}\right)
\end{eqnarray}
assuming that inverse Compton scattering is subdominant for cooling. 
Note that non-thermal emission is characterized by fast cooling regimes ($\gamma'_{\rm{m}}>\gamma'_{\rm c}$) in the fiducial model. 
The cooling timescale for $\gamma_{\rm{m}}'$ is 
\begin{eqnarray}
\label{eq:t_minc}
t_{\rm c} (\gamma_{\rm{m}}')
\sim 3\times 10\,{\rm{s}} ~
\left(\frac{\epsilon_{\rm{B}}}{0.1}\right)^{-1}
\left[\frac{(\gamma_{\rm sf}-1) (4\gamma_{\rm sf}+3)}{1.5}\right]^{-1}
 \nonumber\\
\left(\frac{\rho_{\rm AGN}}{1\times 10^{-16}\,{\rm g~cm^{-3}}}\right)^{-1}
\left(\frac{\gamma_{\rm{m}}'}{40}\right)^{-1}
\left(\frac{\gamma_{\rm sf}'}{1.2}\right)^{-1}. 
\end{eqnarray}
From $t_{\rm c} (\gamma_{\rm{m}}')$, 
the typical shell width of electrons with $\gamma_{\rm{m}}'$ emitting the synchrotron photons 
is approximated as 
$\Delta_{\rm shell}(\gamma_{\rm{m}}')\sim t_{\rm c} (\gamma_{\rm{m}}')v_{\rm FS}\sim 5\times10^{11}\,{\rm cm}~[t_{\rm c}(\gamma_{\rm{m}}')/30~{\rm{s}}] (\beta_{\rm FS}/0.5)$.

The Lorentz factor below which synchrotron self-absorption is effective is \citep{Rybicki1979,Fouka2011}
\begin{align}
\label{eq:gamma_a}
\gamma_{\rm{a}}'=\gamma_{\rm{m}}'\times 
 \left(\tau_{\rm q}C_{q+1}\right)^{1/(q+4)}, 
\end{align}
where 
\begin{eqnarray}
\label{eq:tau_a}
\tau_q=
\frac{\pi}{3\sqrt{3}}
\frac{(q^2+q-2)\gamma_{\rm 1}'^{-5}}{1-(\gamma_{\rm 2}'/\gamma_{\rm 1}')^{-q+1}}
\frac{e 
\rho'_{\rm sf} 
\Delta_{\rm shell}(\gamma_{\rm{a}}') }{B'_{\rm sf}{\rm sin}\theta_{\rm PA} m_{\rm{p}}},
\end{eqnarray}
$\theta_{\rm PA}$ is the typical pitch angle between the magnetic field and the velocity of electrons, $\gamma_2$ and $\gamma_1$ are the maximum and minimum Lorentz factors of non-thermal electrons with power-law index of $q$ 
around $\gamma_{\rm{a}}$ while the self-absorption heats electrons, respectively, 
\begin{eqnarray}
\label{eq:cp}
C_q=\frac{2^{(q+1)/2}}{q+1}
\Gamma\left(\frac{q}{4}-\frac{1}{12}\right)
\Gamma\left(\frac{q}{4}+\frac{19}{12}\right), 
\end{eqnarray}
$\Gamma$ is the Gamma function, 
$C_{p+1}\sim1.6$ for $q=2.5$, 
$n'_{\rm norm}$ is the normalization for the electron number density, 
and we assume that electrons are randomly oriented in the frame of the shocked fluid. 
We set $\gamma_2'=\gamma_{\rm{max}}'$, $\gamma_1'=\gamma_{\rm{m}}'$, and $q=p$ for $\gamma_{\rm{m}}'<\gamma_{\rm{a}}'$, 
and $\gamma_2'=\gamma'_{\rm{m}}$, $\gamma_1'=\gamma_{\rm{a}}'$, and $q=2$ for $\gamma_{\rm{a}}'<\gamma_{\rm{m}}'$. 
We determine $\gamma_{\rm{a}}'$ to satisfy Eqs.~\eqref{eq:gamma_a} and \eqref{eq:tau_a}.
For $\gamma_{\rm{m}}'<\gamma_{\rm{a}}'$, 
we assume that the power-law slope of the Lorentz factor of electrons during synchrotron self-absorption is the same as the one injected ($p$) at around $\gamma'=\gamma_{\rm{a}}'$.
If synchrotron self-absorption becomes effective after the synchrotron cooling modifies the electron distribution to the slope of $p-1$, 
$\gamma_{\rm{a}}'$ is enhanced by a factor of $\sim 1.08$ in the fiducial model 
(Eq.~\ref{eq:gamma_a}).

With the synchrotron frequency given by 
\begin{align}
\label{eq:nu_sync}
\nu_{\rm sync}
&=\gamma_{\rm sf} \frac{e B'_{\rm sf}}{2\pi m_{\rm{e}} c}\nonumber\\
&\simeq 2\times 10^{9}~{\rm Hz}
\left(\frac{\gamma_{\rm sf}}{1.2}\right)
\left(\frac{B'_{\rm sf}}{600~{\rm G}}\right), 
\end{align}
the frequencies corresponding to $\gamma_{\rm{m}}'$, $\gamma_{\rm{max}}'$, and $\gamma_{\rm{a}}'$ are, respectively, \citep{Fan2008}
\begin{eqnarray}
\label{eq:nu_min}
\nu_{\rm{m}}\simeq3\times10^{12}~{\rm Hz} 
\left(\frac{\gamma_{\rm{m}}'}{40}\right)^2
\left(\frac{\nu_{\rm sync}}{ 2\times 10^{9}~{\rm Hz}}\right),
\end{eqnarray}
\begin{align}
\label{eq:nu_max}
\nu_{\rm{max}}&\simeq2\times10^{21}~{\rm Hz} 
\left(\frac{\gamma_{\rm{max}}'}{1\times 10^6}\right)^2 \left(\frac{\nu_{\rm sync}}{2\times 10^{9}~{\rm Hz}}\right)\nonumber\\
&\simeq 2\times 10^{21}~{\rm Hz} 
\left(\frac{\xi}{1}\right)^{-1} 
\left(\frac{\beta_{\rm sf}^2\gamma_{\rm sf}}{0.3}\right),
\end{align}
and 
\begin{eqnarray}
\label{eq:nu_a}
\nu_{\rm{a}}=
7\times 10^{12}
~{\rm Hz} 
\left(\frac{\gamma_{\rm{a}}'}{60}\right)^2 
\left(\frac{\nu_{\rm sync}}{2\times 10^{9}~{\rm Hz}}\right).
\end{eqnarray}

For $\nu_{\rm{a}}<\nu_{\rm{m}}$, the luminosity by synchrotron radiation is approximately given by \citep{Fan2008}
\begin{align}
\nu L(\nu) \sim L_{\rm syn} \times 
\left\{
\begin{array}{cl}
 (\nu/\nu_{\rm{m}})^{\frac{-p+2}{2}} 
 ~~~\mathrm{for}~~\nu_{\rm{m}}<\nu<\nu_{\rm{max}}\\
 (\nu/\nu_{\rm{m}})^{1/2} 
 ~~~\mathrm{for}~~\nu_{\rm{a}}<\nu<\nu_{\rm{m}}\\
(\nu/\nu_{\rm{a}})^{3} (\nu_{\rm{a}}/\nu_{\rm{m}})^{1/2}
 ~~~\mathrm{for}~~\nu_{\rm syn}<\nu<\nu_{\rm{a}}.\\
\end{array}
\right.
\end{align}
On the other hand, 
for $\nu_{\rm{m}}<\nu_{\rm{a}}$, electrons are thermalized below and around $\gamma_{\rm{a}}$. 
We assume that the emission by synchrotron radiation in such cases 
is approximately given by
\begin{align}
\nu L(\nu) 
\sim L_{\rm syn} \times 
\left\{
\begin{array}{cl}
 (\nu/\nu_{\rm{m}})^{\frac{-p+2}{2}} + {\rm exp}(1-(\nu/\nu_{\rm{a}})^{1/2}) \nonumber\\
 ~~~\mathrm{for}~~\nu_{\rm{a}}<\nu<\nu_{\rm{max}}\\
 (\nu/\nu_{\rm{a}})^{3}
 ~~~\mathrm{for}~~\nu<\nu_{\rm{a}}, 
\end{array}
\right.
\end{align}
where the second term for $\nu_{\rm{a}}<\nu<\nu_{\rm{max}}$ and the luminosity for $\nu<\nu_{\rm{a}}$ represent emission from thermalized electrons. 
The order of $\nu_{\rm c}<\nu_{\rm{m}}<\nu_{\rm{a}}$ is astrophysically rare \citep{Gao2013}, while it is often realized in the systems we investigated. This is presumably because the ambient material is dense, and the outflow is characterized by its high energy and compact size. 
The synchrotron-self Compton spectrum of synchrotron photons by thermalized electrons is assumed to be 
\begin{align}
\nu L(\nu) 
\sim L_{\rm syn}Y_{\rm SSC} \nonumber\\
\times 
\left\{
\begin{array}{cl}
 {\rm exp}(1-(\nu/\nu_{\rm a,SSC})^{1/2})
 +(\nu/\nu_{\rm m,SSC})^{(2-p)/2} \nonumber\\
 ~~~\mathrm{for}~~\nu_{\rm a,SSC}<\nu\\
 (\nu/\nu_{\rm a,SSC})^{2} \nonumber\\
 ~~~\mathrm{for}~~\nu<\nu_{\rm a,SSC}, 
\end{array}
\right. 
\end{align}
whose thermal components are derived by considering inverse Compton scattering of synchrotron photons 
by electrons thermalized by synchrotron self-absorption (with the distribution $N_{\rm{e}}\propto \gamma^2 {\rm exp}(-\gamma/\gamma_{\rm{a}})$) \citep{Ghisellini1988,Gao2013}, where $\nu_{\rm a,SSC}\approx (4/3) \gamma_{\rm{a}}'^2\nu_{\rm{a}}$ and $\nu_{\rm m,SSC}\approx (4/3) \gamma_{\rm{m}}'^2 \nu_{\rm{m}}$ are the absorption and minimum frequencies for synchrotron-self Compton. 
Note that since the low-energy range of the synchrotron-self Compton spectrum is determined by the up-scattered photon distribution, the synchrotron-self Compton spectrum is softer than the synchrotron spectrum. 
The spectrum of 2nd order inverse Compton scattering is computed in a similar fashion as that of the synchrotron-self Compton. 
For $\epsilon_{\rm{B}}\ll\epsilon_{\rm{e}}$, 
$Y_{\rm SSC}$ and $Y_{\rm 2ndIC}$ are approximated as 
$Y_{\rm SSC}=(\epsilon_{\rm{e}}/\epsilon_{\rm{B}})^{1/3}$ and $Y_{\rm 2ndIC}=(\epsilon_{\rm{e}}/\epsilon_{\rm{B}})^{2/3}$. 
Note that second order inverse Compton scattering is in the Thomson regime if ${\rm{max}}(\nu_{\rm{m}}\gamma'^3_{\rm{m}},\nu_{\rm{a}}\gamma'^3 _{\rm{a}}) < m_{\rm{e}} c^2/2$ for the fast cooling regime
 \citep{Fan2008}. 
We do not consider the third order inverse Compton scattering since 
it is suppressed by the Klein Nishina effect, 
where in the Klein-Nishina regimes the scattering cross section is reduced compared to Thomson scattering due to quantum electrodynamical corrections \citep{Blumenthal1970}.

Fig.~\ref{fig:prop_fid} shows the dependence of the properties of emission on the distance $R$ from the supermassive black hole. 
It can be seen that this dependence is very strong, which can be explained as follows. 
At smaller radii, the scale height of the disk is lower and the density of the AGN disk is higher. Due to the higher gas density, the accretion rate onto stellar-mass BHs is higher, resulting in a stronger jet power and a higher shock velocity. The delay timescale is also shorter for smaller scale height and higher shock velocity (Eq.~\ref{eq:t_break}). Also, the duration is shorter for a higher AGN density and higher shock velocity (Eq.~\ref{eq:t_bo}). The temperature of the thermal emission is higher for higher shock velocity and AGN density (Eq.~\ref{eq:tem_bo}). These are the reasons that properties of emission strongly depend on the radius.

\subsection{Numerical choices }
\label{sec:numerical_choice_m6}

When we discuss the association of ZTF19abanrhr to GW190521 and 
predict future electromagnetic counterparts 
in Results, 
we use the following values for our model parameters. 
As constrained by \citet{Graham20} and \citet{LIGO20_GW190521}, 
we assume that 
the masses of the merged black hole and the supermassive black hole hosting the AGN are $m=150\,\Msun$ and $M=10^8\,\Msun$, respectively. 
The jet energy conversion efficiency to $\eta_{\rm j}=0.5$ \citep{Tchekhovskoy2010}, 
reflecting the spin magnitude of the merged black hole ($a_{\rm BH,rem}\sim 0.7$ \citep{LIGO20_GW190521}). 
In order to reproduce 
the observed luminosity of the AGN in units of the Eddington Luminosity ($L_{\rm Edd}$) to be 
$\sim0.2$ \citep{Graham20} with a radiation efficiency of $\eta_{\rm rad}=0.1$, 
we set the gas inflow rate from the outer boundary ($R_{\rm out}=10\,{\rm{pc}}$) of the AGN disk to ${\dot M}_{\rm in}=50~{L}_{\rm Edd}/c^2$, 
the angular momentum transfer parameter of the AGN disk to $m_{\rm AM}=0.6$ \citep{Thompson05}, 
and the viscous parameter of the AGN disk to $\alpha_{\rm AGN}=0.1$ \citep{Martin2019}. 
We assume that the opening angle of the injected jet is $\theta_0=0.3$ \citep{Hada2018,Berger2014}, 
and the fraction of the accretion rate onto the black hole (${\dot m}$) over the capture rate to $f_{\rm acc}=15$ as discussed above. 
We adopt optimistic values for 
the fraction of postshock energy carried by the post shock magnetic field 
and that by electrons 
to 
$\epsilon_{\rm{B}}=0.1$ 
and $\epsilon_{\rm{e}}=0.3$ (e.g., \citealt{Panaitescu2001,Frail2005,Sironi_2013}), respectively, 
and the power-law slope for injected electrons accelerated by the first order Fermi process to $p=2.5$.

\section{Results}
\label{sec:properties_breakout}

\subsection{Breakout emission from merging black holes }
\label{sec:observability_merging_BHs}

Breakout emission is expected in association with black hole mergers as follows. 
The jet direction is aligned to the black hole spin direction. 
The black hole spin directions before the mergers tend to be aligned 
perpendicular
to the AGN disk plane. Mergers 
are generally expected to occur with a different 
spin direction 
since the angular momentum directions of merging binaries are predicted to be quasi-randomized \citep{Samsing20} due to frequent binary-single interactions \citep{Tagawa20b_spin} in addition to inhomogeneities in the outer regions of the AGN disk \citep{Tagawa20_ecc}. This 
results in the reorientation of the jets. 
After 
this reorientation, 
as the jet once again collides with unshocked gas, 
shocks emerge and breakout emission is released following the emission of gravitational waves (Fig.~\ref{fig:schematic_solitary_merge}b). We predict the properties and detectability of the breakout emission in this model, applied to possible electromagnetic counterparts of gravitational wave events.

\begin{figure*}\begin{center}\includegraphics[width=150mm]{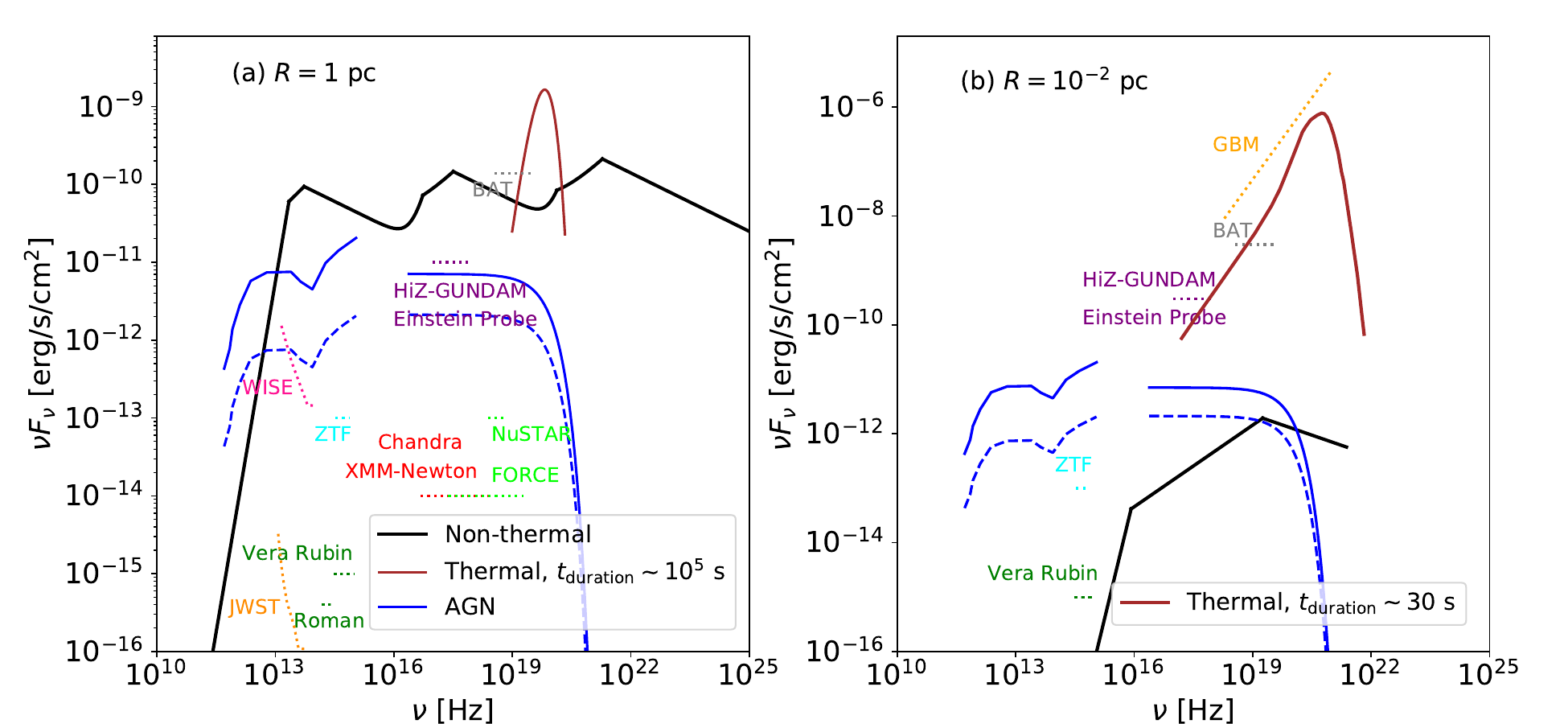}\caption{
The SED for 
non-thermal (solid black) and thermal (solid brown) 
emission from
typical merging black holes in AGN disks 
($m=60~\Msun$ and $d_{\rm L}=500~{\rm Mpc}$) at $R=1~{\rm{pc}}$ with $f_{\rm acc}=15$ (panel~a) 
and at $R=10^{-2}~{\rm{pc}}$ with $f_{\rm acc}=1$ (panel~b). 
Solid and dashed blue lines represent emission from the host AGN and its variability amplitude, respectively. The dotted cyan, dark green, gray, purple, red, light green, pink, and orange lines mark the sensitivity of ZTF, the Vera Rubin Observatory and the Roman space telescope, {\it{Swift}}~BAT, 
HiZ-GUNDAM and Einstein Probe, Chandra and XMM-Newton, NuSTAR and FORCE, WISE, and JWST, respectively. 
The integration time for observations 
is set to 
$t_{\rm duration}$ presented in the legend, 
while $t_{\rm int}=10^4~{\rm{s}}$ 
is adopted for the sensitivities of {\it{Swift}}~BAT in panel~a. 
}\label{fig:l_nu1_m8_flux}\end{center}\end{figure*}

\begin{figure}
\begin{center}
\includegraphics[width=85mm]{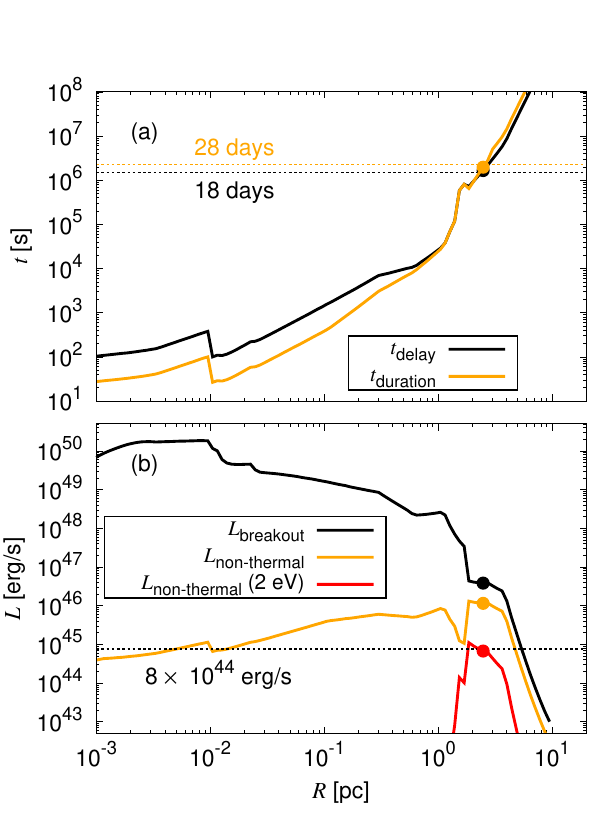}
\caption{
The timescales and luminosity of shock breakout emission produced around merging black holes as a function of the distance from the supermassive black hole ($R$). 
(a): The delay timescale ($t_{\rm delay}$, black) and the observed duration of breakout emission ($t_{\rm duration}$, orange). 
(b): The breakout luminosity of the thermal emission ($L_{\rm breakout}$, black) 
and that by the non-thermal emission at its peak frequency ($L_{\rm non-thermal}$, orange) and at $h\nu=2~{\rm{eV}}$ ($L_{\rm non-thermal}(2~{\rm{eV}})$, red), 
including Doppler beaming. 
The non-thermal emission at $h\nu=2~{\rm{eV}}$ is also shown for the models with lower efficiencies of magnetic field amplification ($\epsilon_{\rm{B}}=0.01$; thin solid blue) and electron acceleration ($\epsilon_{\rm{e}}=0.03$: thin dashed blue). 
The black hole locations adopted in the fiducial model are indicated with filled circles superposed on the lines. 
The dashed holizontal lines present the observed values for $t_{\rm delay}$, $t_{\rm duration}$, and $L$ at $\sim2~{\rm{eV}}$. 
}
\label{fig:prop_fid3}
\end{center}
\end{figure}

\begin{figure*}
\begin{center}
\includegraphics[width=150mm]{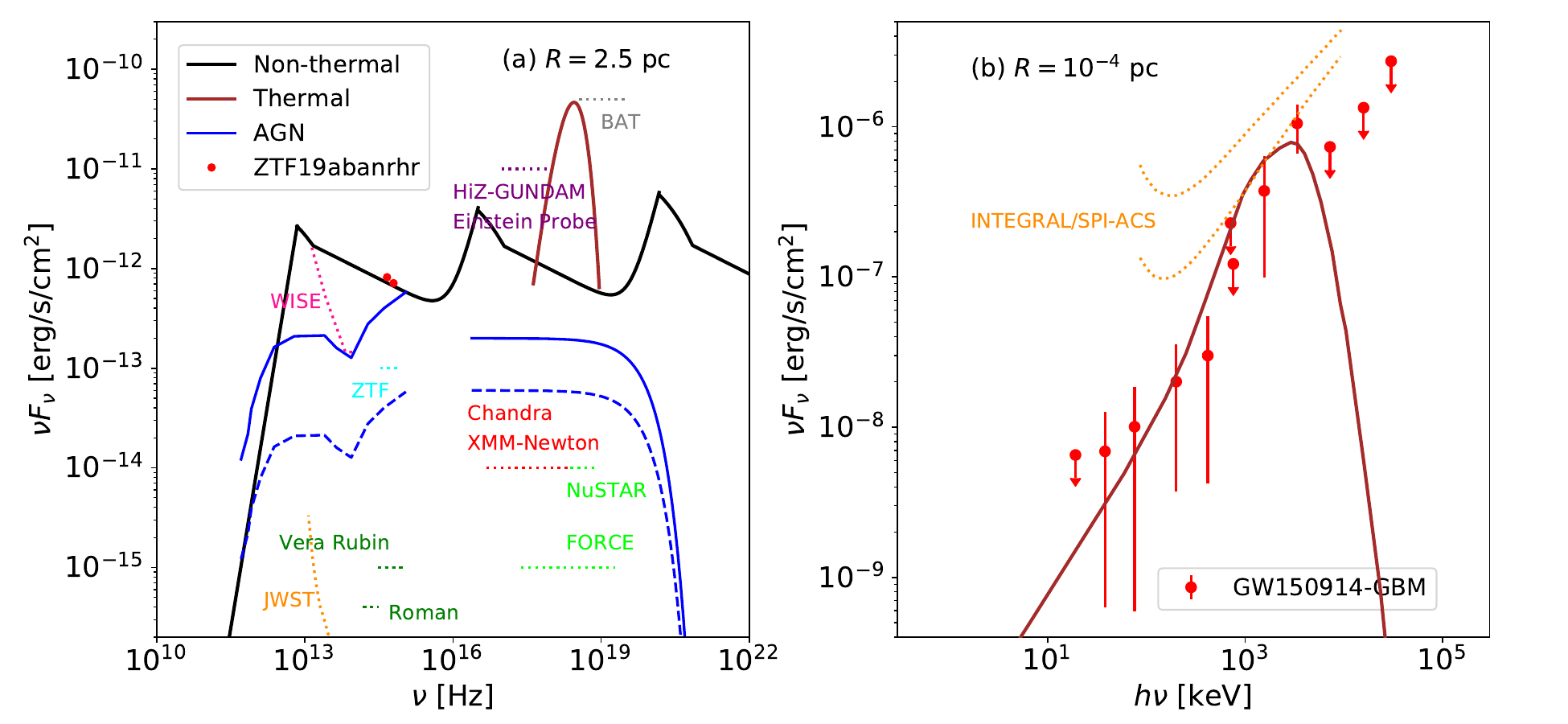}
\caption{
Same as Fig.~\ref{fig:l_nu1_m8_flux}, but the results are specific 
for emission concerning the putative associations with gravitational wave events. 
The SED for 
$R=2.5~{\rm{pc}}$ (panel~a) and $R=10^{-4}~{\rm{pc}}$ (panel~b). 
The dotted cyan, dark green, gray, purple, red, light green, pink, and orange lines mark 
the sensitivity 
of ZTF, the Vera Rubin Observatory 
and the Roman space telescope, 
{\it{Swift}}~BAT (with the integration time of $t_{\rm int}=10^5~{\rm{s}}$), 
HiZ-GUNDAM ($t_{\rm int}=10^4~{\rm{s}}$) and
Einstein Probe ($t_{\rm int}=10^3~{\rm{s}}$), Chandra ($t_{\rm int}=2\times 10^4~{\rm{s}}$) and
XMM-Newton ($t_{\rm int}=10^4~{\rm{s}}$), 
NuSTAR ($t_{\rm int}=10^6~{\rm{s}}$) and FORCE ($t_{\rm int}=10^6~{\rm{s}}$), 
WISE ($t_{\rm int}=10^4~{\rm{s}}$), and JWST ($t_{\rm int}=10^4~{\rm{s}}$), 
respectively. 
The red points in panel~(a) correspond to the observed luminosity of ZTF19abanrhr in the $r$ and $g$ bands assuming $g-r\sim 0.15$, 
and those in panel~(b) represent the observed luminosity of GW150914-GBM with $1$--$\sigma$ error bars. 
The dotted orange curves mark 
the range of sensitivity of {\it INTEGRAL} SPI-ACS assuming $d_{\rm L}=410~{\rm Mpc}$ \citep{Savchenko2016}. 
For thermal emission in panel (b), we use the spectral shape given as the purple line in Fig.~4 in \citet{Ito2020} boosted to a higher energy by $\gamma_{\rm sf,f}=5$. 
}
\label{fig:l_nu1_m8}
\end{center}
\end{figure*}

The SED for emission from
typical merging black holes in AGN disks. 
Same as Fig.~\ref{fig:l_nu1_m8}, 
but the results assume common black hole mergers ($m=60~\Msun$ and $d_{\rm L}=500~{\rm Mpc}$) at $R=1~{\rm{pc}}$ with $f_{\rm acc}=15$ (panel~a) 
and at $R=10^{-2}~{\rm{pc}}$ with $f_{\rm acc}=1$ (panel~b). 
Dotted orange line shows the sensitivity for 
{\it{Fermi}} GBM. 
The integration time is limited by $t_{\rm duration}$ presented in the legend, 
while $t_{\rm int}=10^4~{\rm{s}}$ (panel~a) and $30~{\rm{s}}$ (panel~b) are adopted for the sensitivities of {\it{Swift}}~BAT.

\subsection{Properties of breakout emission }
\label{sec:properties}

We first predict the characteristic properties of the breakout emission from a merging black hole. 
To understand the detectability of the breakout emission from common black hole mergers, 
Fig.~\ref{fig:l_nu1_m8_flux} shows the parameter dependence of 
the spectral energy distribution 
(SED) 
of the breakout emission 
adopting $m=60~\Msun$, $d_{\rm L}=500~{\rm Mpc}$, and $R=1~{\rm{pc}}$ (panel~a) and $R=10^{-2}~{\rm{pc}}$ (panel~b). 
We assume $f_{\rm acc}=15$ for $R=1~{\rm{pc}}$ 
and $f_{\rm acc}=1$ for $R=10^{-2}~{\rm{pc}}$, 
since the accretion rate can be enhanced by the recoil kick if the delay timescale is long enough that shocked gas can accrete onto a black hole before the breakout ($\S$\ref{sec:recoil_accretion}). 
For mergers at $R=1~{\rm{pc}}$, 
non-thermal emission can be 
discovered by ZTF, the Vera Rubin Observatory, 
the Roman space telescope, 
HiZ-GUNDAM, and the Einstein Probe, 
and also be detected by follow-up observations 
by Chandra, XMM-Newton, 
NuSTAR, FORCE, WISE, and JWST once the source direction is localized by ZTF. 
Additionally, thermal emission from $R=1~{\rm{pc}}$ and $R=10^{-2}~{\rm{pc}}$ 
can be detected by the {\it{Swift}}~Burst Alert Telescope (BAT). 
However, for $R=10^{-2}~{\rm{pc}}$, the detection probability is reduced by the beaming effect by a factor of $\sim 1/\gamma_{\rm sf,f}^2\sim 1/16$ compared to cases with $\gamma_{\rm sf,f}\sim1$, 
since the opening angle of the jet at breakout of $\theta_j\sim0.08(<\theta_0)$ \citep{Bromberg2011} is smaller than $1/\gamma_{\rm sf,f}$, 
where $\gamma_{\rm sf,f}$ is the final Lorentz factor of the shocked gas. 
In addition, 
{\it{Swift}} BAT needs to direct to the source beforehand since there is not enough time for the telescope to redirect due to the short delay time ($\sim 10^2~{\rm{s}}$) and duration ($\sim30~{\rm{s}}$), which further reduces the observational probability by a factor of $\sim 10$. 
Thus, 
for mergers in $R\gtrsim{\rm{pc}}$, optical surveys and/or future X-ray monitors can detect electromagnetic counterparts, as suggested for ZTF19abanrhr, and then, infrared and X-ray pointing facilities can detect it later. 
For $R\lesssim 10^{-2}~{\rm{pc}}$, electromagnetic counterparts are difficult to discover unless 
they are bright enough to be detected by the {\it{Fermi}}~GBM and/or the {\it{INTEGRAL}}~SPI-ACS as suggested for GW150914-GBM and LVT151012-GBM.

In Fig.~\ref{fig:l_nu1_m8_flux}~a, 
the luminosity from the host AGN in the relevant energy range is given by 
\begin{align}
\nu L_{\rm AGN}(\nu)\sim 10^{44}~{\rm erg/s}(M/10^8~\Msun)
({\dot M}c^2/L_{\rm Edd})(f_{\rm cor}/10)^{-1}\,.
\end{align} 
where $f_{\rm cor}$ is the correction fraction from the total luminosity to the luminosity at some frequency. 
As depicted by the solid blue lines in 
Figs.~3 and 4, 
we assume that 
$f_{\rm cor}\sim 5$ at $c/\nu=4400~$\AA~ and extrapolate the luminosity 
for $10^{12}~{\rm Hz}\lesssim \nu \lesssim 10^{15}~{\rm Hz}$ using the cyan or blue points in Fig.~7 of \citet{Ho2008} depending on the assumed Eddington ratio, 
and $f_{\rm cor}\sim 10$ in $0.1~{\rm keV}\leq h\nu$ \citep{Ho2008,Trakhtenbrot2017,Duras2020} with the upper exponential cut off at $300~{\rm keV}$ \citep{Ricci2018}. 
We also assume that 
the fraction of the variable luminosity compared to the average luminosity 
in optical bands with $t_{\rm duration}\lesssim 0.1~{\rm yr}$ is $f_{\rm var}\lesssim 0.1$ \citep{Kozlowski2016} 
and that in X-ray bands is $f_{\rm var}\sim 0.3$ (\citealt{Soldi2014,Maughan2019}, dashed blue lines).

\subsection{ZTF19abanrhr and GW190521 }
\label{sec:transients_GW190521}

\subsubsection{Observed properties }

\label{sec:properties_ZTF}

We overview observed properties for ZTF19abanrhr. 
\citet{Graham20} reported that the optical transient, ZTF19abanrhr, from the AGN ${\rm J124942.3+344929}$ at $z=0.438$ is possibly associated with GW190521. 
The optical flare in ZTF19abanrhr began to exceed the persistent flux from the AGN activity after $\sim 18$ days (in the rest frame) from the occurrence of the merger event, GW190521. 
The peak luminosity of ZTF19abanrhr is $\sim 8\times 10^{44}\,{\rm erg/s}$ in the $g$ and $r$ bands with the observed duration of $\sim 28\,{\rm days}$ (in the rest frame). 
Before and after the peak, 
the luminosity increases and decreases roughly exponentially. 
The change in the slope of the luminosity is somewhat shallower in the latter. The color is consistent with a constant with $g-r\sim 0.2$, while it appears to be reduced to $g-r\sim 0.15$ during the flare. 
The mass of the supermassive black hole hosting the AGN ${\rm J124942.3+344929}$ is $M\sim 10^8$--$10^9\,\Msun$, and 
the luminosity of the AGN in units of the Eddington luminosity is $\sim 0.02$--$0.2$ \citep{Graham20}. 
The mass of the merged remnant is 
$m\sim 150\,\Msun$ and its dimensionless spin is $a_{\rm BH,rem} \sim 0.7$ \citep{LIGO20_GW190521}. 
The inclination angle (angle between the total angular momentum of the merged binary with respect to the line-of-sight) is estimated to be $\sim 40$--$60^{\circ}$ \citep{LIGO20_GW190521,CalderonBustillo21}, disfavoring a transient with high luminosity due to a significant beaming of emission from the Blandford-Znajek jet or its shock, since the spin of a merged remnant, and accordingly the Blandford-Znajek jet, are expected to be aligned with the total angular momentum of the merged binary. Note that the inclination angle estimated by \citet{Gayathri2022} is different from that above. Thus, the possible beaming of emission associated with gravitational wave events may be useful to constrain the parameters of black hole mergers observed by gravitational waves.

The association significance of ZTF19abanrhr to GW190521 is under debate \citep{Ashton2020,Palmese2021,CalderonBustillo21}. 
\citet{Ashton2020} analyzed that a chance occurrence probability of the association of GW190521 and ZTF19abanrhr is between $\sim 8\%$--$ 50\%$. \citet{Palmese2021} estimated that the chance occurrence probability is $\sim 4\%$ and $\sim 70\%$ by using a damped random walk model adopted in \citet{Graham20} and a more general structure function, respectively. Here, a damped random walk model well describes structures of flares from AGN disks for the duration of flares to be less than $\sim {\rm yr}$ \citep{Kozlowski2016}. \citet{CalderonBustillo21} found that with different priors (uniform in inverse of a mass ratio, $1$--$4$, instead of uniform in a mass ratio, $1/4$--$1$), the chance occurrence probability is reduced to $\sim 1\%$, due to the modification for the position of GW190521 in different preferences for the mass ratio. Currently, the association is considered to be tentative. 
There are also several independent suggestions that GW190521 originated in an AGN disk \citep{Yang20_gap,Tagawa20_MassGap}, especially if the suggested high eccentricity is true \citep{Tagawa20_ecc,Samsing20,Romero-Shaw20,Gayathri2022}.

\subsubsection{Model for ZTF19abanrhr}

\label{sec:ztf_model}

We here discuss whether the possible association of ZTF19abanrhr to GW190521 
can be explained by emission from black holes merging in an AGN disk. 
We focus on this association here, but note that other optical associations were reported \citep{Graham2022} after the present manuscript was being prepared for submission. 
The properties of the additional flares are similar to those of ZTF19abanrhr and broadly consistent with our model. 
The fits in our model to these flares require a range of different parameter choices, and will be presented in a follow-up paper. 
We adopt parameter values to match the observed properties of this source ($\S$\ref{sec:numerical_choice_m6}). 
To propose a consistent model for this association, 
it is crucial to simultaneously explain (I) 
all of the several observed 
properties of the flare 
and (II) the reason why there is no bright emission before the merger. 

(I) Several properties of the flare are obtained by the ZTF observations, which include its luminosity in the optical bands, its color, and its time evolution ($\S$\ref{sec:properties_ZTF}). 
Fig.~\ref{fig:prop_fid3} 
shows the properties of 
thermal and non-thermal emission at the shock breakout of a jet produced from merging black holes. 
The discontinuities at distances from the central supermassive black hole of $R\sim1~{\rm{pc}}$ and $\sim0.01~{\rm{pc}}$ are, respectively, due to the transition from sub-relativistic to relativistic regimes and to gap formation around black holes by their gravitational torques 
caused by the transition for the opacity of AGN-disk gas. 
In this model, the delay timescale ($t_{\rm delay}\sim19~{\rm days}$, Fig.~\ref{fig:prop_fid3}a) and the observed duration of the breakout emission for non-thermal photons ($t_{\rm duration}\sim23~{\rm days}$, panel~a) are, respectively, comparable to the observed delay time and the duration (dotted gray lines), provided the merger occurs at $R\sim2.5$ pc. 
In the breakout emission, the delay timescale is calculated by the time for the shock to reach the edge of the AGN disk (Eq.~\ref{eq:t_break}), while the duration is calculated by the time that photons diffuse out from a breakout shell to the edge of the AGN disk (Eq.~\ref{eq:t_bo}, see $\S$\ref{sec:breakout}). 
Since the observable properties especially the timescales are mostly influenced by $R$ (Fig.~\ref{fig:prop_fid3}), flares similar to previous one (ZTF19abanrhr if it was real) are expected to be observed in the future if there is a hot spot for the merging location ($\S$\ref{sec:discussions}). 
Fig.~\ref{fig:l_nu1_m8} shows the spectral energy distribution (SED) of the emission at $R=2.5~{\rm{pc}}$. 
The luminosity at $\sim6\times10^{14}~{\rm Hz}$, 
arising from synchrotron emission from 
the non-thermal electrons accelerated at the forward shock, 
is roughly consistent with the observed value of $\nu L_{\nu}\sim8\times10^{44}~{\rm erg/s}$. 
The power-law spectral slope ($s$, $\nu L_{\nu} \propto \nu^{s}$) 
of the synchrotron emission 
is given by 
$s=-0.25$, 
while the observed slope during the flare 
corresponding to the color $g-r=0.15$--$0.2$ is $s=-(0.45$--$0.6)$ \citep{Graham20}. Since the contribution of the flare to the background AGN luminosity is $\sim30\%$ and the slope of the background emission is $s\sim-0.6$ ($g-r=0.2$), 
the slope for the combination of the flare and the background emission is estimated to be $\sim-0.5$ at the observed wavelengths of $~470~{\rm nm}$ and $\sim 650~{\rm nm}$. 
Hence the synchrotron emission 
is roughly consistent with the observed color during the flare. 
The remarkable point of the estimates is that the delay time, the duration, and the luminosity of ZTF19abanrhr are well reproduced for the same value of the location within the disk, that is
$R=2.5~{\rm{pc}}$ (Fig.~\ref{fig:prop_fid3}), although the luminosity is somewhat adjusted ($\S$\ref{sec:numerical_choice_m6}). 
For the non-thermal emission, 
the peak of the lightcurve likely comes when the forward shock reaches the edge of the AGN disk \citep{Perna2021_GRBs}. 
A more quantitative estimate of the lightcurves for the non-thermal emission will require numerical simulations, which is beyond the scope of this paper (see $\S$\ref{sec:synchrotron}).

(II)
Another issue, which needs to be resolved to claim the association, 
is that intense emission should be launched only after the merger of GW190521. 
In the models for emission from the shock around the jet, 
this requires (a) the existence of cold gas in the direction of the jet propagation 
from the merger remnant and (b) gas accretion within the delay between GW190521 and ZTF19abanrhr. 
The fact that the emission is produced only after the merger has 
not been
addressed or fully explained 
in previous models 
for the association. Recoil kicks might offer a solution. 
\citet{Graham20} proposed that the merger remnant moves to unperturbed dense gas as a result of the recoil kick and begins to accrete. However, it is unclear why circumbinary gas should not be present and power bright emission already prior to the merger \citep{Farris2015,Tang2018}. 
One possibility is that a cavity pre-exists around the remnant, carved out by radiative and/or mechanical feedback from the merging black holes; the merger remnant then moves to unshocked regions at the boundary of this cavity by recoil kicks \citep{Kimura2021_BubblesBHMs}. 
On the other hand, with the parameters ($R\gtrsim {\rm{pc}}$) constrained above, 
it takes too long ($\gtrsim {\rm yr}$)
to reach replenished gas, compared to the $\sim$ month delay between GW190521 and ZTF19abanrhr ($\S$\ref{sec:probability}). 
As discussed above, the model for breakout emission can explain why bright emission is observed only after the mergers (Fig.~\ref{fig:schematic_solitary_merge}b). 
We conclude that breakout emission from shocks driven by a Blandford-Znajek jet can explain the properties of the transient, ZTF19abanrhr, possibly associated with GW190521.

We further predict how the breakout emission similar to ZTF19abanrhr can be observed by various telescopes in the future. 
We employ parameters 
to explain ZTF19abanrhr as a fiducial model (see $\S$\ref{sec:numerical_choice_m6}). 
Fig.~\ref{fig:l_nu1_m8} shows the SED for the breakout emission assuming a luminosity distance $d_{\rm L}\sim 3~{\rm Gpc}$ to the event together with the sensitivity curves of various equipments. 
Non-thermal emission 
can be detected by ZTF, the Vera Rubin Observatory, 
and the Roman space telescope
(solid black, dotted cyan, and dotted green lines in Fig.~\ref{fig:l_nu1_m8}; see $\S$\ref{sec:telescopes} for their properties). 
Thermal emission can be detected 
by 
future wide-field X-ray surveys, HiZ-GUNDAM and the Einstein Probe 
if breakout emission is produced closer to us (e.g. $d_{\rm L}\lesssim{\rm Gpc}$). 
Chandra, XMM-Newton, 
Nuclear Spectroscopic Telescope Array (NuSTAR), 
Focusing On Relativistic universe and Cosmic Evolution (FORCE, a future hard X-ray telescope), 
Wide-field Infrared Survey Explorer (WISE), and the James Webb Space Telescope (JWST) can detect 
emission from such merger remnants, if the merger event 
is well localized by other observations, since the time spent for directing to the source ($\sim {\rm day}$) is shorter than the duration of the emission for mergers at $R\gtrsim {\rm{pc}}$ ($\sim 0.1~{\rm yr}$, Fig.~\ref{fig:prop_fid3}b). 
Since the timing and the duration of the emission is the same 
at all wavelengths, from infrared to gamma-rays, the 
emission can be 
simultaneously observed in a wide range of wavelengths 
with bright optical emission. 
We propose that such follow-up observations in infrared and X-ray bands can be a smoking gun signature of this scenario, leading to a first confirmation of the origin of the black hole mergers. 
Then, we can further derive properties of the mergers' environments as described for possible several associations above and in $\S$\ref{sec:method}.

\subsection{GW150914-GBM and LVT151012-GBM }
\label{sec:transients_GW150914}

\subsubsection{Observed properties}
\label{sec:properties_GW150914}

We next 
describe the observed properties of GW150914-GBM. 
Associated with GW150914, 
the {\it Fermi} GBM might have detected a transient of luminosity $\sim 2\times 10^{49}\,{\rm erg/s}$ at energies $\sim 10~{\rm keV}$--several ${\rm MeV}$ 
with spectral shape 
$\nu L_{\nu}\propto \nu^{p_{\rm{s}}}$ with $p_{\rm{s}}\sim 0.6$. 
The possible transient was observed $\sim0.4\,{\rm{s}}$ after the gravitational wave event with duration of $\sim 1~$s.
The signal-to-noise ratio was estimated to be 5.1 with a false-alarm probability for the association with GW150914 of 0.0022 (2.9$\sigma$) \citep{Connaughton2016}. 
Several criticisms and/or issues were raised by a number of studies \citep{Greiner2016,Savchenko2016}, 
and they are answered or discussed in \citet{Connaughton2018}. 
A possible problematic point is that
the anti-coincidence shield (ACS) of the Spectrometer on board {\it INTEGRAL} (SPI) 
put constraints on the 
gamma-ray intensities in $75~{\rm keV}$--$2~{\rm MeV}$ in the direction of GW150914 at the merger \citep{Savchenko2016}. 
On the other hand, since {\it INTEGRAL} is most sensitive at $\sim 100~{\rm keV}$ (Fig.~3 in \citet{Savchenko2016}), while the peak energy inferred for GW150914-GBM is $\sim 3.5^{+2.1}_{-1.1}~{\rm MeV}$ and the estimated spectral index is hard, it can be barely consistent with the no detection by {\it INTEGRAL} SPI-ACS \citep{Connaughton2018}
(see Fig.~3b). 
\citet{Connaughton2018} discussed that a weaker signal estimated by the analyses in \citet{Greiner2016} may be preferred for the consistency with the no detection by {\it INTEGRAL} SPI-ACS.

Additionally, 
\citet{Bagoly2016} reported another possible electromagnetic association with the gravitational wave event, LVT151012, with false alarm probability of $0.037$. 
LVT151012 is a tentative gravitational wave event for the merger of $23^{+18}_{-5}~\Msun$ and $13^{+4}_{-5}~\Msun$ at $d_{L}=1100^{+500}_{-500}~{\rm Mpc}$ with 
a false alarm rate of $0.44~{\rm yr}^{-1}$ and a false alarm probability of $0.02$ \citep{Abbott2016_TwoGWs}. 
The burst looks similar to GW150914-GBM. It has a similar flux, it occurred within $\sim ~{\rm{s}}$ from the gravitational wave event with duration of $\sim {\rm{s}}$, and the peak energy is found to be between $130~{\rm keV}$ and $3.5~{\rm MeV}$.

Although, unlike for GW190521, the inclination angle (angle between the total angular momentum of the merged binary with respect to the line-of-sight) for GW150914 and LVT151012 was not significantly constrained due to the lack of detection of the higher-order multipole modes \citep{Abbott2016_GW150914_prop}, constraints on the inclination angle in future GW observations associated with gamma-ray counterparts would be very useful, since low values of the inclination angle are required to explain the association of GW150914-GBM by our model as discussed in the next section.

\subsubsection{Model for GBM events}

We also find 
that the properties, including the luminosity, delay time, duration and color of the transients suggested to be associated with GW150914 and LVT151012 can be explained by thermal emission from black hole mergers at $R=10^{-4}~{\rm{pc}}$ and $R\lesssim 10^{-4}~{\rm{pc}}$, respectively.

A model for this candidate source needs to explain the observed hard spectrum (red circles in Fig.~\ref{fig:l_nu1_m8}b). 
A possible process reproducing the spectrum is 
thermal emission. 
The peak energy of the transient is about $\sim 3~{\rm MeV}$, 
while annihilation of gamma rays prohibits strong thermal emission above $\sim{\rm MeV}$ in the rest frame \citep{Budnik2010,Ito2020}. 
To explain the strong emission at $\sim3~{\rm MeV}$, 
the observer needs to be within 
the beaming direction of the shock at breakout with 
a Doppler shift by a factor of $\sim4$--$10$ (Fig.~\ref{fig:l_nu1_m8}b). 
The spectral shape of the thermal emission in Fig.~\ref{fig:l_nu1_m8}b is adopted from that calculated by \citet{Ito2020} considering $\gamma\gamma$ annihilation. 
For these events, 
we assume the Bondi-Hoyle-Lyttleton rate for the accretion rate onto the merged remnant ($\S$\ref{sec:accretion_remnants}).

We constrain model parameters as follows. 
First, to produce the spectral peak energy, we adopt $\gamma_{\rm sf,f}\sim5$ reflecting the discussion above. 
The observed delay timescale of $t_{\rm delay}\sim0.4~{\rm{s}}$ 
and $\gamma_{\rm sf,f}\sim5$ together can be used to constrain
the scale height of the AGN disk to $H_{\rm AGN}\sim 3\times 10^{11}~{\rm cm}(\gamma_{\rm sf,f}/5)^2 (t_{\rm delay}/0.4~{\rm{s}})$ (see Eq.~\ref{eq:t_break_rela}).
The observed breakout luminosity ($L_{\rm breakout}$) 
is proportional to the jet power ($L_{\rm j}$) and the accretion rate onto the black hole ${\dot m}$ (Eq.~\ref{eq:lj_macc}), which is adjusted by changing the AGN density at the location of the black hole ($\rho_{\rm AGN}$) and the supermassive black hole mass ($M$) given 
the mass of the merged remnant 
($m$), $R$, and $\gamma_{\rm sf,f}$ \citep{Tagawa2022_BHFeedback}. 
Additionally, 
the initial opening angle of the jet $\theta_0$ at launch can be determined to reproduce $\gamma_{\rm sf,f}$ with given $H_{\rm AGN}$, $\rho_{\rm AGN}$, and $L_{\rm j}$ \citep{Bromberg2011}. 
By adopting $\rho_{\rm AGN}=4\times10^{-6}~{\rm g/cm^3}$, $M=10^6\,\Msun$, $R=10^{-4}~{\rm{pc}}$, and $\gamma_{\rm sf,f}=5$, 
we can derive $\theta_0=0.06$, $H_{\rm AGN}\sim3\times 10^{11}$, $L_{\rm j}=10^{48}~{\rm erg/s}$, $L_{\rm breakout}=1.5\times10^{49}~{\rm erg/s}$ (including Doppler beaming), the delay time $t_{\rm delay}=0.4~{\rm{s}}$ (using Eq.~\ref{eq:t_break_rela}), and the duration $t_{\rm duration}=0.2~{\rm{s}}$ (using Eq.~\ref{eq:t_ang}). 
Note that $t_{\rm duration}=0.2~{\rm{s}}$ is shorter than the inferred total duration of $\sim1.0~{\rm{s}}$. On the other hand, the duration of the phase with the above bright luminosity is $\lesssim1/4~{\rm{s}}$ (Fig.~7 in \citet{Connaughton2016}), consistent with the value derived here (see also \citet{Bagoly2016}). 
The light-curve is expected to grow super-exponentially at breakout \citep{Sapir2011}, and after breakout emission, the luminosity decays as a power-law in time \citep{Nakar2012}. The inferred density ($\rho_{\rm AGN}=4\times10^{-6}~{\rm g/cm^3}$) together with $R=10^{-4}~{\rm{pc}}$, $H_{\rm AGN}\sim 3\times10^{11}$, and $M=10^6\,\Msun$ is realized for an AGN disk with a common accretion rate 
of ${\dot M}\sim0.2~{L}_{\rm Edd}/c^2$ 
and a viscous parameter of $\alpha_{\rm AGN}=0.1$. 
Thus, even without identifying a host AGN, if our model correctly predicts the electromagnetic emission, 
we can derive properties of the AGN, such as $H_{\rm AGN}$, $\rho_{\rm AGN}$, and $R$, where black hole mergers often occur. 
The spectral energy distribution (SED) 
adjusted to explain GW150914-GBM is presented in Fig.~\ref{fig:l_nu1_m8}b.

Although we consider the collision of the jet with the inner regions of the AGN disk, it may also be possible to produce similar emission by considering the collision of the jet with a circum-black hole disk, 
whose size is $\sim3\times10^{11}~{\rm cm}$ for 
the mass of the merged remnant 
of $m\sim60~\Msun$ and the accretion rate onto the black hole of ${\dot m}\sim10^4~l_{\rm Edd}/c^2$ 
assuming a slim disk model \citep{Abramowicz1988}.

The properties of the transient possibly associated with LVT151012 ($\S$\ref{sec:properties_GW150914}) can also be explained by a similar parameter set, while the higher luminosity (due to the larger distance by a factor of $\sim 2$) and the lower peak energy (roughly around MeV) require a higher $\rho_{\rm AGN}$ and a lower $\gamma_{\rm sf,f}$. 
These can be explained by a merger at smaller $R$, where $\rho_{\rm AGN}$ is higher and a lower $\gamma_{\rm sf,f}$ is preferred to reproduce the delay time ($\sim{\rm{s}}$) due to a smaller $H_{\rm AGN}$.

In summary, in the model of emission from merging black holes in an AGN disk, 
there exists a set of parameters for which the properties of ZTF19abanrhr, GW150914-GBM, and LVT151012-GBM can be explained.

\section{Discussions}
\label{sec:discussions}

In this section, we discuss 
the probability of observing breakout emission from merging black holes, 
the possible obscuration by a dust torus, 
the parameter dependence of the results, 
and the merging location within the AGN disc.
Specifications of several telescopes which might be able to detect the breakout emission at different wavelengths are summarized in Appendix~\ref{sec:telescopes}.

\subsection{Probability of observing breakout emission from merging black holes }
\label{sec:probability}

We suggest that breakout emission from merging black holes may be easier to discover compared to that from solitary black holes 
due to the following reasons: 
(A) By using gravitational wave observations, constraints on the spatial localization and the timing of flares are 
modestly and significantly improved, respectively, thus removing the need to continuously monitor large sky areas to search for rare flares. 
(B) The accretion rates onto merger remnants may be enhanced due to recoil kicks (see below). 
(C) The spin magnitudes of black holes are enhanced to $a_{\rm BH,rem}\sim0.7$ by the merger \citep{Buonanno08}, increasing the jet luminosity. Although the spin magnitudes of solitary black holes ($a_{\rm BH,iso}$) are highly uncertain, spin-up by accretion and spin-down by the Blandford-Znajek jet may be roughly equal at around $a_{\rm BH,iso}\lesssim0.3$ (e.g., Fig.~10 of \citealt{Narayan2021}). Assuming $a_{\rm BH,iso}\sim0.3$, the jet luminosity is enhanced by a factor of $\sim (a_{\rm BH,rem}/a_{\rm BH,iso})^2 \sim5(a_{\rm BH,rem}/0.7)^{2}(a_{\rm BH,iso}/0.3)^{-2}$. 
(D) The merging black holes may be more massive than isolated black holes as inferred from gravitational wave observations \citep{Abbott21_GWTC3}, producing brighter emission due to higher accretion rates.

In order for 
emission to be produced shortly after mergers (roughly within a timescale comparable to the duration of the flare) and 
 be discovered using gravitational wave observations, there are several key elements which are needed.
First, (a) gas needs to be present along the direction of propagation of the jet (black hole spin) of the merged remnant, or otherwise thermal and non-thermal emission from dissipation of the jet power would not be expected. 
The shape of the ejected region is cylindrical 
with an aspect ratio 
of $\sim\theta_{\rm 0}f_{\rm ext}\sim1(\theta_0/0.3) (f_{\rm ext}/3)$ \citep{Tagawa2022_BHFeedback}, where 
$\theta_0$ is the opening angle of the injected jet, 
$f_{\rm ext}$ is 
the fraction of the radial extent of the shocked regions advanced after the breakout from the AGN disk over that at the breakout, 
and $f_{\rm ext}\sim3$ 
is motivated for a spherical explosion in a background with an exponential gas density profile \citep{Olano2009}. 
Hence, 
gas presumably exists at the merger 
except in the cylindrical regions that have been ejected before the merger in previous episodes of jets launched from the progenitor black holes. 
Since the angular momentum directions of binaries are predicted to be random at mergers in AGN disks due to frequent binary-single interactions \citep{Tagawa20b_spin,Samsing20} 
and inhomogeneities in the outer regions of the AGN disk, 
the directions of the spins of the merged remnants and that of the jet propagation are plausibly randomized at mergers. 
Thus, gas often exists in the direction of the spin of the merged remnants 
(Fig.~\ref{fig:schematic_solitary_merge}b), 
with a probability of $P_{\rm col}\sim{\rm cos}({\rm atan}(\theta_{\rm 0}f_{\rm ext}))\sim0.7$ for $\theta_{\rm 0}\sim0.3$ and $f_{\rm ext}\sim3$. 
If we assume $\theta_0\sim0.03$--$0.5$ \citep{Berger2014,Hada2018} and $f_{\rm ext}\sim 1$--$5$ \citep{Olano2009}, $P_{\rm col}$ ranges in the interval $\sim0.4$--$1$.

The second issue is that 
(b) gas needs to be accreted onto the merged remnant for a flare to occur in association with a black hole binary merger. Here, we assume that 
the flare can be considered to be associated with a gravitational wave event 
when the delay timescale is roughly comparable to or less than the duration of the flare, 
since it is probably difficult to find an association with a short duration and a long delay time. 
In \citet{Kimura2021_BubblesBHMs}, it is proposed that electromagnetic counterparts may be produced by moving to unshocked regions by the gravitational wave recoil kick after merger. 
Usually, 
gas replenishment by the recoil kick takes a longer time compared to $t_{\rm delay}$ and $t_{\rm duration}$. 
The kick velocity due to gravitational wave radiation ($v_{\rm rk}$) is on the order of $\sim10^2$--$10^3~{\rm{km/s}}$ and gas is depleted within $r_{\rm dep}\sim~f_{\rm ext}H_{\rm AGN}\theta_{\rm 0}\sim 10^{16}\,{\rm cm} (R/{\rm{pc}})$ \citep{Tagawa2022_BHFeedback}, where $H_{\rm AGN}$ is the scale height of the AGN disk at $R$. Then, the kicked black hole crosses the cavity on the timescale of 
\begin{align}
\label{eq:tarr}
t_{\rm arr}\sim r_{\rm dep}/v_{\rm rk}\sim30\,{\rm yr}~(v_{\rm rk}/3\times 10^2\,{\rm{km/s}})^{-1}(R/3~{\rm{pc}}). 
\end{align}
Such a long delay time ($t_{\rm arr}$) compared to the short duration 
(orange line in Fig.~2~a) 
would not be regarded as association in observations. 
In ZTF19abanrhr, the observed delay time of $\sim 0.05~{\rm yr}$ is much shorter than the arrival timescale in Eq.~\eqref{eq:tarr}. 
To expect a flare with a short delay 
whose timescale is comparable to the observed delay time, 
the merging black holes need to have circum-black hole disks at the merger. 
Adopting the model in \citet{Tagawa2022_BHFeedback}, 
we derive that 
$P_{\rm active}\sim0.03$ at $R\sim3~{\rm{pc}}$, 
where $P_{\rm active}$ is the 
probability that a merged remnant accompanies a circum-black hole disk and a jet,
although the active probability 
strongly depends on model parameters. 
On the other hand, 
if mergers are significantly facilitated by circum-black hole disks, which is not assumed in the discussions above, 
merged remnants may tend to accompany circum-black hole disks \citep{Farris2015,Tang2018}. 
\citet{Bartos17} assessed mergers to be significantly facilitated by circum-black hole disks, while \citet{Tagawa19} found the opposite by applying the updated theory of gap formation \citep{Duffell14}. Note that both studies assumed that the accretion onto black holes is limited at around the Eddington rate, which may underestimate the effect on binary evolution by circum-black hole disks 
because the disk mass can be much higher for highly super-Eddington accretion disks; 
hence the influence of torques from circum-black hole disks on mergers has not been quantitatively understood. 
Since $P_{\rm active}$ is estimated to be larger than $\sim 0.01$ \citep{Tagawa2022_BHFeedback}, and significant facilitation of mergers by circum-binary disks may enhance $P_{\rm active}$ up to 1, we assume as a rough estimate that $P_{\rm active}$ ranges within $\sim 0.01$--$1$. 

Third, (c) 
the breakout emission 
needs to be 
bright enough 
to overshine the host AGN emission 
and 
(d) bright enough to 
be detected by current (and future) facilities. 
However, the probabilities that 
the breakout emission is brighter than the host AGN emission ($P_{\rm bright}$) 
and that can be detected by facilities ($P_{\rm det}$) 
are 
uncertain and difficult to constrain theoretically as the luminosity is influenced by several uncertain parameters (such as $\epsilon_{\rm{B}}$, $\epsilon_{\rm{e}}$, $f_{\rm acc}$, and $\rho_{\rm AGN}$) 
and also transients with a high shock Lorentz factor ($\gamma_{\rm sf,f}$), as suggested for GW150914-GBM, are rarer by a factor of ($\sim \gamma_{\rm sf,f}^2$) to be in the beaming direction of the shock. 
On the other hand, 
GW emission for on-axis events is stronger, which 
enhances the detection rate of GWs associated with electromagnetic emission beamed in our direction, assuming that the direction of the jet (and the spin of the merger remnant) is perpendicular to the orbital plane of merging binary.

Finally, (e) black hole mergers need to occur in AGN disks. 
Although this probability ($P_{\rm AGN}$) has been highly debated \citep{Gayathri2021_AGN_O3}, it has been less constrained and ranges from 0 to 1.

Considering these factors 
and adopting the fiducial values for $P_{\rm col}$ and $P_{\rm active}$ and $P_{\rm bright}P_{\rm det}P_{\rm AGN}=1$, 
the probability for observing breakout emission associated with mergers in AGN disks is $P_{\rm association}\sim~P_{\rm col}P_{\rm active}P_{\rm bright}P_{\rm det}P_{\rm AGN}\sim0.02(P_{\rm col}/0.7)(P_{\rm active}/0.03)(P_{\rm bright}P_{\rm det}P_{\rm AGN}/1)$. 
According to this estimate, 
if 
$1$--$9$ transients 
possibly including seven optical \citep{Graham2022} and two gamma-ray flares \citep{Connaughton2016,Bagoly2016}
are actually associated with the mergers among O(90) mergers discovered by LIGO/Virgo/KAGRA, as modeled above, $P_{\rm bright}P_{\rm det}P_{\rm AGN}$ can be constrained to be high ($\gtrsim0.5$). 
Note that there are significant uncertainties as $P_{\rm col}P_{\rm active}$ ranges from $4\times 10^{-3}$ to $1$, and $P_{\rm bright}$, $P_{\rm det}$, and $P_{\rm AGN}$ range from 0 to 1 as discussed above. 
Nevertheless, we expect that future observations, such as by the Vera Rubin Observatory 
and the Roman space telescope
will significantly 
increase $P_{\rm det}$ 
especially for dimmer breakout emission from less bright AGNs 
as well as 
constrain $P_{\rm col}P_{\rm active}$, $P_{\rm bright}$, and $P_{\rm AGN}$ 
due to their high sensitivity and wide field-of-view.

\subsection{Possible obscuration by a dust torus }
\label{sec:dust_torus}

If abundant dust exists along the line of sight to the observer, 
emission in the optical to soft X-ray bands can be absorbed. 
Due to the existence of a dust torus around AGNs, 
the breakout emission in these bands may hence be largely absorbed. 
In particular, our model predicts a merger at $R\sim 2.5~{\rm{pc}}$ for ZTF19abanrhr, which is beyond the dust sublimation radius 
($\sim 0.1~{\rm{pc}}$ for the AGN luminosity of $\sim 10^{44}~{\rm erg/s}$ in the ultraviolet bands, \citealt{Barvainis1987}), 
suggesting the existence of dust in this region. 
On the other hand, for $R\sim$a few pc, 
dust in regions above and below the AGN disk plane is possibly cleared out by AGN disk winds \citep{Wada2016} and/or supernovae feedback \citep{Wada2009} (see also observations, \citealt{Stalevski2017}). 
If this is the case, 
dust in the disk is usually confined to the disk plane and 
can be evaporated locally by the cocoon feedback, 
as also confirmed in gamma-ray bursts \citep{Waxman2000}. 
The breakout emission can therefore emerge without obscuration 
by dust in this case. 
Conversely, measured properties (e.g. normalization and reddening) of the breakout emission from $R\sim $a few pc may be useful to understand the geometry of dust tori. 
Also, if the breakout emission is produced from black holes 
in type II AGNs unlike the host of ZTF19abanrhr, the emission in optical and soft X-ray bands is likely obscured. 
Since the emission is unobscured in infrared and hard X-ray bands even if the emission is produced in geometrically thick dust tori, 
simultaneous observations with infrared/hard X-ray and optical bands will be able to constrain the configuration of dust tori. Thus, wide-field infrared/hard X-ray facilities, such as {\it Swift}~BAT, will play important roles to find the electromagnetic counterparts produced in dust tori.
Additionally, due to the obscuration, $P_{\rm obs}$ is reduced 
by the covering fraction of the dust torus. 
Although the influence of dust is uncertain, 
our model can avoid the issue of dust obscuration to explain ZTF19abanrhr 
for the reasons discussed above.

\begin{figure*}
\begin{center}
\includegraphics[width=160mm]{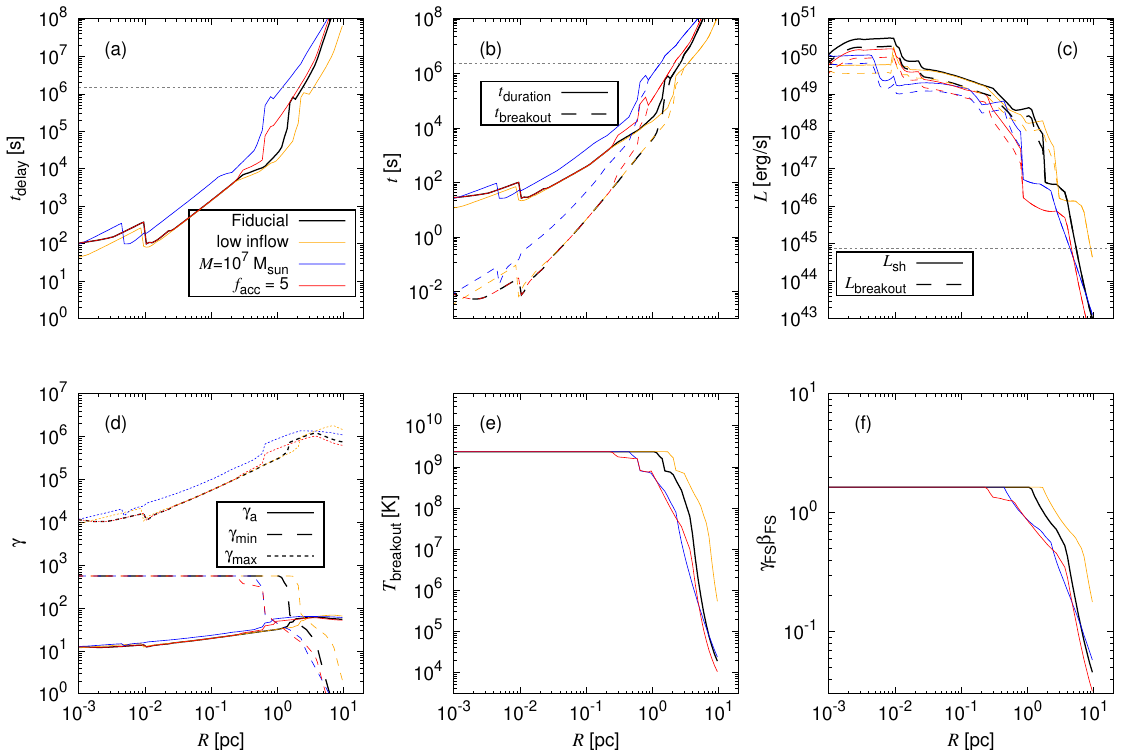}
\caption{
Similar to 
Fig.~\ref{fig:prop_fid} 
but results for several choices of the parameters are shown. 
Black, orange, blue, and red lines, respectively, present results for 
the fiducial model, and the models with 
${\dot M}_{\rm in}=2~\Msun/{\rm yr}$, 
$M=10^7~\Msun$ and ${\dot M}_{\rm in}=1~\Msun/{\rm yr}$, 
and $f_{\rm acc}=5$. 
(a)~The delay time. 
(b)~The duration (solid) and the breakout timescale (dashed). 
(c)~The shock (solid) and breakout (dashed) luminosities. 
(d)~The absorption (solid), minimum (dashed), and maximum (dotted) Lorentz factors for non-thermal electrons. 
(e)~The breakout temperature. 
(f)~The dimensionless forward shock velocity ($\gamma_{\rm FS}\beta_{\rm FS}$). 
}
\label{fig:prop_dep}
\end{center}
\end{figure*}

\subsection{Parameter dependence }
\label{app_sec:parameter_dependence}

Fig.~\ref{fig:prop_dep} shows the parameter dependence of the properties of breakout emission, as shown in 
Fig.~\ref{fig:prop_fid} 
for the fiducial model. 
For low ${\dot M}_{\rm in}$ (orange and blue lines) or low $f_{\rm acc}$ (red lines), 
the accretion rate onto the BH is rather low, 
which lowers $L_{\rm j}$, $L_{\rm sh}$, and $L_{\rm breakout}$ (panel~c). 
At large $R$, for low ${\dot M}_{\rm in}$ (orange lines), 
the luminosity is higher than that of the fiducial model. This is because 
for low ${\dot M}_{\rm in}$ 
the AGN disk becomes geometrically thin in outer regions, 
in which the jet head is less decelerated before the breakout 
because the jet sweeps up a lower amount of gas. This 
enables a higher speed for the shock. 
Note that the AGN disk density in outer regimes, where the Toomre parameter satisfies $\sim 1$, only depends on the orbital frequency, and is unaffected by ${\dot M}_{\rm in}$ (see Eq.~3 of \citealt{Thompson05}). 
Although there is the exception as explained above, to produce bright emission (high $L_{\rm sh}$), high ${\dot M}_{\rm in}$ and/or high $f_{\rm acc}$ are indeed required.

The fiducial model parameters and their influence on properties of the breakout emission are listed in Table~\ref{table:parameter_fiducial}. 
There are several sensitive parameters, $R$, $f_{\rm acc}$, $\epsilon_{\rm{B}}$, and $\epsilon_{\rm{e}}$. 
$L_{\rm break}$ is significantly influenced by all of the four parameters, while $t_{\rm delay}$ and $t_{\rm duration}$ are mostly influenced by $R$. 
According to our model, $R$ can be constrained from the timescales, while $f_{\rm acc}$, $\epsilon_{\rm{B}}$, and $\epsilon_{\rm{e}}$ can be constrained if non-thermal emission is observed 
at multiple wavelengths. 

\begin{table*}
\begin{center}
\caption{
Fiducial values of our model parameters and their influence on properties of the breakout emission. 
For "sensitivity", 
we made a rough classification of the parameters influencing the results ($L_{\rm breakout}$, $t_{\rm delay}$, $t_{\rm duration}$) by 
more than two orders of magnitude ("very sensitive"), 
one order-of-magnitude ("sensitive"), 
a factor of $\gtrsim 2$ ("moderate"), 
and a factor of $\lesssim 2$ ("insensitive").
We also add the comment "constrained" for parameters which are well constrained by the observations of ZTF19abanrhr. 
}
\label{table:parameter_fiducial}
\hspace{-5mm}
\begin{tabular}{p{6.5cm}|p{2.5cm}|p{6.5cm}}
\hline 
Parameter & Fiducial value &Sensitivity to results \\
\hline\hline
Radial distance of the black hole from the supermassive black hole & $R=2.5\,{\rm{pc}}$& very sensitive 
on the three properties\\\hline
Ratio of the black hole accretion rate to the gas capture rate & $f_{\rm acc}=15$&sensitive
on $L_{\rm breakout}$ and moderate on $t_{\rm delay}$ and $t_{\rm duration}$
\\\hline
Conversion efficiency of shock kinetic energy to magnetic/electron energy & $\epsilon_{\rm{B}}=0.1$, $\epsilon_{\rm{e}}=0.3$&very sensitive 
on reducing $L_{\rm non-thermal}$\\\hline
Jet energy conversion efficiency & $\eta_{\rm j}=0.5$& sensitive on $L_{\rm breakout}$ and moderate on $t_{\rm delay}$ and $t_{\rm duration}$
but constrained\\\hline
Mass of the merged remnant 
& $m=150\,{\Msun}$& sensitive on $L_{\rm breakout}$ and moderate on $t_{\rm delay}$ and $t_{\rm duration}$
but constrained\\\hline
Gas inflow rate from the outer boundary of the AGN disk 
& ${\dot M}_{\rm in}=50~{L}_{\rm Edd}/c^2$&sensitive on the three properties but constrained\\\hline
Mass of the supermassive black hole & $M=10^8\,{\Msun}$& sensitive on the three properties but constrained\\\hline
Opening angle at the base of the jet & $\theta_{\rm 0}=0.3$&moderate on the three parameters\\\hline
Angular momentum transfer parameter in outer regions of the AGN disk & $m_{\rm AM}=0.5$& moderate on the three parameters\\\hline
Viscous parameter of the AGN disk & $\alpha_{\rm AGN}=0.1$&insensitive on the three parameters\\\hline
\end{tabular}
\end{center}
\end{table*}

\subsection{Merging location }
\label{sec:location}

We have proposed 
that black hole mergers at $R\gtrsim{\rm{pc}}$, $R\sim10^{-4}\,{\rm{pc}}$, 
and $R\lesssim10^{-4}~{\rm{pc}}$ can explain the properties of the transients possibly associated with GW190521, GW150914, and LVT151012, respectively. 
Here we discuss whether such diversity in the positions of mergers in AGN disks is expected. 
Mergers have been predicted to be concentrated in gap-forming regions \citep{Tagawa19,Tagawa20_MassGap} and migration traps \citep{McKernan19,Secunda18,Yang20_gap} in the inner AGN disk, or in the outer regions \citep{Perna2021_AICs} of the disk (but may also occur throughout the disk, \citealt{McKernan19}), 
which have similar influence on the evolution and mergers of black hole binaries. 
Gaps are predicted to form for massive black holes or in a thin disk \citep{Tagawa19},
while migration traps may form at 
$\lesssim10^{3}~R_{\rm g} 
\sim10^{-4}~{\rm{pc}}~(M/2\times10^6~\Msun)$ \citep{Bellovary16} (but see \citealt{Pan2021_MT}), 
where $R_{\rm g}=GM/c^2$, $G$ is the gravitational constant, and $c$ is the speed of light. 
If heating torques are taken into account for the dynamics of gas around black holes \citep{Hankla2020}, they can reverse the direction of migration at various distances from the supermassive black hole, while the evolution may be further complicated due to shocked bubbles \citep{Tagawa2022_BHFeedback}. 
Also, the migration timescale without gaps is longer in outer regions especially around massive supermassive black holes 
 \citep{Pan2021_MT}, 
in which mergers are also expected 
at $R\gtrsim{\rm{pc}}$ \citep{Perna2021_AICs}. 
As ZTF19abanrhr is believed to be associated with an AGN powered by a supermassive black hole of $M\sim 10^8\,\Msun$, 
inefficient migration (possibly due to long migration timescale, gaps, or heating torques) may be consistent with mergers at $R\gtrsim{\rm{pc}}$. 
Also, the mergers at $R\lesssim10^{-4}\,{\rm{pc}}$, suggested for the transients possibly associated with GW150914 and LVT151012, 
may be consistent with mergers in migration traps of AGN disks around less massive supermassive black holes with $M\sim10^5$--$10^6\,\Msun$. 
Conversely, if these transients are actually associated with merging black holes in AGN disks, this places constraints on
the positions of gap-forming regions and/or migration traps, 
and yields
information on the AGN disks from the properties of the transients.

Since transients are reported for the massive mergers (including GW190521) or the early events (GW150914 and LVT151012), the detection of transients may require mergers of massive black holes and/or mergers at close distances from us. 
Indeed, the breakout luminosity ($L_{\rm breakout}$), the jet power ($L_{\rm j}$), and the accretion rate onto a black hole (${\dot m}$) depend on 
the mass of the merged remnant 
($m$) 
as $L_{\rm breakout}\propto~L_{\rm j}\propto{\dot m}\propto m^{2/3}$ 
(Eq.~1 of \citet{Tagawa2022_BHFeedback}), 
and the flux is higher for closer events. 
Also, for transients with a longer delay ($t_{\rm delay}$), such as ZTF19abanrhr, enhancement of the accretion rate due to recoil kicks ($\S$\ref{sec:recoil_accretion}) can be expected for massive remnants since the circum-black hole disk mass within some radius $r$ from the black hole is $m_{\rm CBD}(<r)\propto~m^{11/5}$ \citep{Haiman2009}.

\section{Summary}

In this paper, we investigate properties of the EM emission associated with black hole mergers in AGN disks. 
Both thermal and non-thermal radiation is produced by shocks, through collisions between the AGN disc gas and 
a Blandford-Znajek jet powered by the merger-remnant BH, in a direction reoriented at the merger. 
We suggest that the emission emerging as the jets break out from the optically thick AGN disc can self-consistently account for several features of the tentative EM counterparts for the three GW events 
ZTF19abanrhr, GW150914-GBM, and LVT151012-GBM. In particular, the model accounts for the luminosity in the optical bands, 
as well as the timing and duration 
of the emission. 
Our main results are summarized as follows. 

\begin{enumerate}

\item 
For mergers at a few pc from the central SMBH, the properties of ZTF19abanrhr are reproduced by non-thermal emission. 
In this event, the accretion rate onto the black hole is predicted to be enhanced by shocks due to recoil kicks. 

\item
For mergers in closer vicinity of the central SMBH, the properties of GW150914-GBM and LVT151012 are reproduced by thermal emission. 

\item
The variety of distances from the central SMBH at which mergers are predicted to take place is consistent with predictions of models for the AGN disk-embedded BH population, in which the BHs migrate slowly and produce frequent mergers. 

\end{enumerate}

To confirm this scenario, 
discoveries of electromagnetic counterparts for future GW events 
are highly desired. 
Then, a smoking-gun signature, 
i.e. bright infrared and X-ray emission concurrent with the optical transient, can be detected by 
WISE, JWST, Chandra, XMM-Newton, 
NuSTAR, and
FORCE, 
once host galaxies are localized by optical observations, 
and/or {\it Swift}~BAT could be able to detect the breakout emission. 
If the scenario is confirmed, these systems will be highly useful to improve our understanding of 
the evolution of compact objects in AGN disks, the structure of AGN disks, 
plasma physics, 
and the expansion history of the universe by helping to constrain the Hubble constant, in addition to unveiling the origin of black hole mergers.

\acknowledgments

This work was financially supported 
by Japan Society for the Promotion of Science (JSPS) KAKENHI 
grant Number JP21J00794 (HT) and 22K14028 (S.S.K.). S.S.K. acknowledges the support by the Tohoku Initiative for Fostering Global Researchers for Interdisciplinary Sciences (TI-FRIS) of MEXT's Strategic Professional Development Program for Young Researchers.
Z.H. was supported by NASA grant NNX15AB19G and NSF grants AST-2006176 and AST-1715661.
R.P. acknowledges support by NSF award AST-2006839.
I.B. acknowledges the support of the Alfred P. Sloan Foundation and NSF grants PHY-1911796 and PHY-2110060.

\appendix

In Appendix, 
we describe the notations, 
the enhancement of the accretion rate due to recoil kicks, 
and the properties of telescopes.

\vspace{\baselineskip}

\section{Notation}
\label{sec:notation}

The notations of variables are listed in Table~\ref{table_notation}.

\begin{table*}
	\caption{
		Notation. 
 	}
\label{table_notation}
\hspace{-0.0mm}
\begin{tabular}{p{2.5cm}|p{5cm}||p{2.5cm}|p{5cm}}
\hline
Symbol&Description&Symbol&Description\\\hline
$r$& 
The distance from the stellar-mass black hole 
&
$H_{\rm AGN}$& The scale height of the AGN disk 
\\\hline

$r_{\rm{b}}$, $r_{\rm ub}$&
The radius within which all gas in a circum-black hole disk is bound and 
that beyond which all gas is unbound after the recoil kick
& 
$r_{\rm acc}$& 
The radius at which gas accretes onto the stellar-mass black hole on the breakout timescale
\\\hline

$R_{\rm g}$& The gravitational radius of a supermassive black hole $GM/c^2$
&
$r_{\rm k}$&
The gravitational radius after a recoil kick $Gm/v_{\rm rk}^2$
\\\hline

$d_{\rm L}$, $d_{\rm L,3Gpc}$&
The luminosity distance to the source and that in unit of $3~{\rm Gpc}$
& 
$\Delta_{\rm shell}(\gamma)$&The shell width of electrons with $\gamma$ emitting non-thermal photons
\\\hline

$d_{\rm edge}$, $d_{\rm edge,BO}$&
The distance to the surface of the AGN disk and that for the breakout emission
& 
$R_{\rm out}$&The outer boundary of the AGN disk
\\\hline

$m$& 
The mass of the merged remnant black hole 
&
$m_{\rm CBD}(r)$ &The mass of the circum-black hole disk within $r$ from a stellar-mass black hole
\\\hline

${\dot m}$, ${\dot M}$ & The accretion rate onto a stellar-mass and a supermassive black hole, respectively&
${\rm{m}}_{\rm BHL}$& The Bondi-Hoyle-Lyttleton rate
\\\hline

${\dot m}_{\rm Edd}$, ${\dot M}_{\rm Edd}$&
The Eddington accretion rate onto a stellar-mass black hole and a supermassive black hole 
& $l_{\rm Edd}$, $L_{\rm Edd}$ &
The Eddington luminosity of a stellar-mass and a supermassive black hole
\\\hline

$L$, $\nu L_{\nu}$& The luminosity and the luminosity at the frequency $\nu$
&
$L_{\rm j}$& The kinetic luminosity of the jet 
\\\hline

$L_{\rm sh}$ & 
The kinematic power of the shock in the jet head 
&$L_{\rm breakout}$, $L_{\rm non-thermal}$ & The breakout luminosity by thermal emission and
that by the non-thermal emission at its peak frequency, respectively
\\\hline

${\tilde L}$ & The ratio between the energy density of the jet and the rest-mass energy density of the surrounding medium 
&$L_{\rm syn}$ & The luminosity of synchrotron emission
\\\hline

$\eta_{\rm j}$ & The conversion efficiency of mass to the jet
&
$\eta_{\rm rad}$& The conversion efficiency of mass to radiation in an accretion disk
\\\hline

$Y_{\rm SSC}$, $Y_{\rm 2ndIC}$ &The powers of synchrotron-self Compton and second order inverse Compton scattering compared to that of synchrotron emission
&$E_{\gamma, \rm BO}$ & The energy of a photon
\\\hline

$t_{\rm duration}$, $t_{\rm delay}$& The duration and delay time of a transient
&$t_{\rm break}$,$t_{\rm break,rel}$, $t_{\rm GW, break}$ &The breakout timescale of a non-relativistic and relativistic shock and gravitational wave, respectively
\\\hline

$t_{\rm diff}$& The diffusion timescale of photons out of the disk
&
$t_{\rm ang}$&The angular timescale
\\\hline

$t_{\rm dyn,CBD}(r)$&The dynamical timescale of a circum-black hole disk at $r$
&$t_{\rm vis}(r)$, $t_{\rm vis,sh}(r)$&
The viscous timescale of a circum-black hole disk at $r$ before and after a shock arises due to a recoil kick
\\\hline

$t_{\rm int}$& The integration time of observations
&
$t_{\rm arr}$& The timescale for a black hole arriving outside a cavity 
\\\hline

$t_{\rm sph}$& The time of transition between the planar and spherical geometries of the breakout shell
&
$t_{\rm dyn}$& The dynamical timescale of a shell
\\\hline

$t_{\rm trans}$& The transparent timescale
&
$T_{\rm breakout}$
&The breakout temperature
\\\hline

$\gamma_{\rm FS}$, $\gamma_{\rm sf}$, $\beta_{\rm FS}$, $\beta_{\rm sf}$&
The Lorentz factor of the forward shock and shocked fluid, their dimensionless velocity
&
$\gamma_{\rm FS,f}$, $\gamma_{\rm sf,f}$, $\beta_{\rm FS,f}$, $\beta_{\rm sf,f}$&
The final Lorentz factor of the forward shock and shocked fluid, their dimensionless velocity
\\\hline

$\gamma_{\rm{e}}$&The Lorentz factor of a electron
&
$\gamma_{\rm{m}}$, $\gamma_{\rm{max}}$, 
$\gamma_{\rm{a}}$, $\gamma_{\rm c}$&
The minimum, maximum, absorption, and cooling 
Lorentz factors
\\\hline

\end{tabular}
\end{table*}

\setcounter{table}{1}
\begin{table*}
	\caption{continued
 	}
\label{table_notation7}
\hspace{-0.0mm}
\begin{tabular}{p{2.5cm}|p{5cm}||p{2.5cm}|p{5cm}}
\hline
Symbol&Description&Symbol&Description\\\hline

$\nu$& The frequency of a photon& 
$\nu_{\rm{syn}}$, $\nu_{\rm{m}}$, $\nu_{\rm{max}}$, 
$\nu_{\rm{a}}$,$\nu_{\rm{m,SSC}}$, $\nu_{\rm{a,SSC}}$ 
&
The synchrotron frequency for electrons with $\gamma_{\rm{e}}=1$, $\gamma_{\rm{m}}$, $\gamma_{\rm{max}}$, 
and $\gamma_{\rm{a}}$, and synchrotron self-Compton frequency for $\gamma_{\rm{m}}$ and $\gamma_{\rm{a}}$
\\\hline

$a_{\rm BH}$&
The dimensionless spin parameter of a black hole&
$a_{\rm BH,rem}$, $a_{\rm BH,iso}$& 
The dimensionless spin parameter of a merged remnant and an isolated black hole
\\\hline

$f_{\rm{acc}}={\dot m}/{\dot m}_{\rm BHL}$& 
The enhancement factor of the accretion rate of the captured gas&
$f_{\rm ext}$& The fraction that the cocoon proceeds to the $r$-direction after the breakout 
\\\hline

$P_{\rm col}$& 
The probability that a jet launched from a merged remnant collide with unshocked gas&
$P_{\rm active}$& The probability that a merged remnant is accreting immediately after merger
\\\hline

$P_{\rm obs}$& 
The probability that the emission from the breakout of the shock from the AGN disk is observable
&
$v_{\rm rk}$& The recoil velocity due to anisotropic emission of gravitational waves at merger
\\\hline

$f_{\rm{b}}$, $f_{\rm ub}$& The factors for recoil kicks $f_{\rm{b}}=r_{\rm{b}}/r_{\rm k}$, $f_{\rm ub}=r_{\rm ub}/r_{\rm k}$
The fraction that the breakout emission is observable
&
$f_{\rm accum}$&The fraction of gas within
$r_{\rm ub}$ accumulated within $r_{\rm{b}}$ after the recoil kick 
\\\hline

$f_{\rm inc}$& The factor that the gas mass within $r_{\rm{b}}$ is enhanced by the recoil shock
&$f_{\rm beaming}$& The beaming factor taking into account an angular effect
\\\hline

$f_{\rm corr}$& The correction factor from the total luminosity to the luminosity at some frequency
&$f_{\rm var}$& The ratio of the variable luminosity compared to the average luminosity at some frequency
\\\hline

$P_{\rm association}$& The probability that black hole mergers in AGN disks accompany electromagnetic observations
&
$i$& The inclination angle between the jet and the orbital angular momentum of the AGN disk 
\\\hline

$\theta_{\rm j}$, $\theta_{\rm 0}$& The jet opening angle and that at injection
&$\theta_{\rm rk}$& The angle of the recoil kick with respect to the plane of the circum-black hole disk
\\\hline

$\theta_{\rm c}$& The opening angle of the cocoon
&
$\tau$&
The optical depth of the AGN disk
\\\hline

$\kappa_{\rm CBD}$, $\kappa_{\rm AGN}$& The opacity of the circum-black hole disk and the AGN disk 
&
$\alpha_{\rm CBD}$
& The viscous parameter for the circum-black hole disk 
\\\hline

$\rho_{\rm AGN}$, $n_{\rm AGN}$& The density and the number density of the AGN disk &
$\rho_{\rm sf}$& The density of the shocked material
\\\hline

$e_{\rm sf}$&The energy density of the shocked material
&
$B_{\rm sf}$& The magnetic field of the shocked material
\\\hline

$\xi$&The parameter representing the ratio of the mean free path to the Larmor radius of electrons
&
$\tau_q$, $q$, $\gamma_1$, $\gamma_2$, $n_{\rm norm}$, $C_q$
&
The parameters for the optical depth by synchrotron self-absorption
\\\hline

$\theta_{\rm PA}$&The pitch angle between the magnetic field and the velocity of electrons
&$p$&
The power-law slope for injected electrons accelerated by the first order Fermi process 
\\\hline

$N(\gamma)d\gamma$&The number of electrons at $\gamma_e=\gamma\pm d\gamma/2$
&
$p_{\rm dm}$&
The power-law slope for the mass profile of the circum-black hole disk
\\\hline

$s$&
The power law slope for electromagnetic radiation at some frequency&
$e$&The elementary charge
\\\hline

$c$ & The speed of light & 
$G$& Gravitational constant
\\\hline

$k_{\rm{B}}$& The Boltzmann constant
&
$h$& The Planck constant
\\\hline

$m_p$& The proton mass& 
$a$& The radiation constant
\\\hline

\end{tabular}
\end{table*}

\section{Enhancement of the accretion rate onto merger remnants}
\label{sec:recoil_accretion}

For accretion onto merged remnants, 
the accretion rate can be significantly affected by 
the recoil kicks due to anisotropic radiation of gravitational wave. 
We follow \citet{Rossi2010} for estimating the enhancement of the accretion rate below. 
After the recoil kick, 
all gas in an accretion disk keeps bound within the radius of 
\begin{align}
\label{eq:rb}
r_{\rm{b}}&
=f_{\rm{b}}r_{\rm k}=
\frac{f_{\rm{b}}Gm}{v_{\rm rk}^2}\nonumber\\
&=5\times 10^{12}~{\rm cm} 
\left(\frac{f_{\rm{b}}}{0.2}\right)
\left(\frac{m}{150~\Msun}\right)
\left(\frac{v_{\rm rk}}{300~{\rm{km/s}}}\right)^{-2}, 
\end{align}
while 
all gas is unbound beyond the radius of 
\begin{align}\label{eq:rub}r_{\rm ub}=f_{\rm ub}r_{\rm k},\end{align}
where $r_{\rm k}\equiv Gm/v_{\rm rk}^2$, 
$f_{\rm{b}}=[-{\rm cos}\theta_{\rm rk}+({\rm cos}\theta_{\rm rk}^2+1)^{1/2}]^2$ and 
$f_{\rm ub}=[{\rm cos}\theta_{\rm rk}+({\rm cos}\theta_{\rm rk}^2+1)^{1/2}]^2$ 
are the factors related to the the kick direction 
(e.g. $f_{\rm{b}}\sim 0.2$ and 
$f_{\rm ub}\sim 5$ for $\theta_{\rm rk}=30^{\circ}$), 
and 
$\theta_{\rm rk}$ is the angle of the recoil kick with respect to the plane of the circum-black hole disk. 
After the recoil, the accretion onto the black hole is expected only from the circum-black hole disk within the radius $r_{\rm ub}$. 
For gas initially within $r_{\rm ub}$, 
its angular momentum is rearranged by the kick 
and its energy is lost due to shocks within the circum-black hole disk. 
The fraction of gas within $r_{\rm ub}$ accumulated within $\sim r_{\rm{b}}$ after the recoil kick ($f_{\rm accum}$) is 
found to be $f_{\rm accum}\sim 20\%$--$50\%$. 
Here, $\sim 20\%$ is roughly derived from the enhancement of the surface density within $r_{\rm{b}}$ in Fig.~19 of \citet{Rossi2010}, while $\sim 50\%$ is roughly derived from the statement 
that all bound gas falls within $r_{\rm{b}}$. 
Note that in \citet{Rossi2010}, a black hole is treated as a sink particle with a size of $0.5~r_{\rm{b}}$, within which gas particles are removed from the simulation (which is described in \citet{Ponce2012}), 
which presumably significantly mitigates the surface density in Fig.~19 of \citet{Rossi2010}. 
Additionally, \citet{Rossi2010} 
adopted 
a more centrally concentrated mass 
profile of $m_{\rm CBD}(<r) \propto r^{p_{\rm dm}}$ with $p_{\rm dm}=1/2$, 
while 
in the gas pressure dominated region with a fixed opacity, the profile is less centrally concentrated, with 
$p_{\rm dm}=7/5$ 
 \citep{Haiman2009}. 
In the case of accretion disks, 
the factor ($f_{\rm inc}$) by which
the gas mass and the gas density 
within $r_{\rm{b}}$ is enhanced by the recoil shock 
is roughly estimated as 
\begin{align}
f_{\rm inc}&\sim 
\frac{f_{\rm accum}[m_{\rm CBD}(<r_{\rm ub})-m_{\rm CBD}(<r_{\rm{b}})]}
{m_{\rm CBD}(<r_{\rm{b}})}\nonumber\\
&\sim f_{\rm accum}(r_{\rm ub}/r_{\rm{b}})^{p_{\rm dm}}, 
\end{align}
which is up to $\sim 60 (f_{\rm accum}/0.5)$ depending on $\theta_{\rm rk}$ 
for $p_{\rm dm}=7/5$.

Hence, 
it is numerically confirmed that the accretion rate can be enhanced by the recoil kick in 
the following steps. 
First, the gas in the annulus between $r_{\rm{b}}$ and $r_{\rm ub}$ is perturbed by the recoil kick and it immediately experiences shocks. Due to the shock, the gas loses its angular momentum, it migrates inward on the dynamical timescale, and then, the gas density in the inner regions of the circum black hole disk and the accretion rate onto the black hole are enhanced.

Without the recoil kick, the accretion mass within $t_{\rm break}$ is 
$m_{\rm CBD}(<r_{\rm acc})$, where $r_{\rm acc}$ is the radius at which gas accretes onto the black hole on the breakout timescale, $t_{\rm break}$ ($t_{\rm vis}(r_{\rm acc})=t_{\rm break}$, where $t_{\rm vis}(r)$ is the viscous timescale of the circum-black hole disk at the distance $r$ from the black hole), and is estimated as 
\begin{align}
\label{eq:t_vis}
r_{\rm acc}
=5\times 10^{12}~{\rm cm}
\left(\frac{t_{\rm break}}{0.1~{\rm yr}}\right)^{5/7}
\left(\frac{m}{150~\Msun}\right)^{1/7}\nonumber\\
\left(\frac{
{\dot m}c^2/{l}_{\rm Edd}
}{8\times 10^4}\right)^{2/7} 
\left(\frac{\alpha_{\rm CBD}}{0.1}\right)^{4/7}
\left(\frac{\kappa_{\rm CBD}}{0.4~{\rm cm^2/g}}\right)^{1/7}
\end{align}
for the standard disk 
in the gas pressure dominated regime \citep{Haiman2009}, 
which is expected at $r=r_{\rm acc}$ for the fiducial model but for $f_{\rm acc}=1$ 
(inner regions are estimated to follow a slim disk model \citep{Abramowicz1988}, while the disk at $r=r_{\rm acc}$ follows the standard disk model), 
where 
$l_{\rm Edd}$ is the Eddington luminosity of the stellar-mass black hole, 
$\kappa_{\rm CBD}$ is the opacity of the circum-black hole disk, 
$\alpha_{\rm CBD}$ is the alpha parameter for the standard disk, 
and 
${\dot m}c^2/{l}_{\rm Edd}=8\times 10^4$ 
is the value in the fiducial model 
with $f_{\rm acc}=1$ at $R=2.5~{\rm{pc}}$. 
After the recoil kick, the accretion mass within $t_{\rm break}$ is 
\begin{align}
\label{eq:f_inc}
\sim f_{\rm inc} m_{\rm CBD}(<r_{\rm{b}})
\frac{t_{\rm break}}{{\rm{max}}[t_{\rm break}, t_{\rm vis,sh}(r_{\rm{b}})]}, 
\end{align}
where $t_{\rm vis,sh}(r)$ is the viscosity of the circum-black hole disk at $r$ after the recoil shock occur in the circum-black hole disk, which can be shorter than $t_{\rm vis}(r)$ due to the shock heating and the possible transition to the slim disk \citep{Abramowicz1988}. 

When the breakout timescale ($t_{\rm break}$) is shorter than $t_{\rm vis,sh}(r_{\rm{b}})$, 
which is satisfied with $v_{\rm rk}\lesssim 300~{\rm{km/s}}$ for $m=150~\Msun$ (Eqs.\ref{eq:rb} and \ref{eq:t_vis}), 
the recoil kick does not eject gas which can accrete onto the black hole within the breakout timescale. In this case, 
using Eq.~\eqref{eq:f_inc}, 
the accretion rate and accordingly the averaged jet luminosity can be enhanced due to the recoil shock by a factor of 
\begin{eqnarray}
&\sim f_{\rm inc}\frac{m_{\rm CBD}(<r_{\rm{b}})t_{\rm vis}(r_{\rm acc})}{m_{\rm CBD}(<r_{\rm acc})t_{\rm vis,sh}(r_{\rm{b}})}\nonumber\\
&\sim f_{\rm inc}\frac{t_{\rm vis}(r_{\rm{b}})}{t_{\rm vis,sh}(r_{\rm{b}})} \gtrsim f_{\rm inc} 
\end{eqnarray}
where we use the rough relation of $m_{\rm CBD}(<r_{\rm{b}})/t_{\rm vis}(r_{\rm{b}})\sim m_{\rm CBD}(<r_{\rm acc})/t_{\rm vis}(r_{\rm acc})$. 
In addition, the enhancement of accretion demands that the breakout timescale be longer than the dynamical timescale within which the shocked gas at $r=r_{\rm{b}}$ can accrete onto the black hole, $t_{\rm dyn,CBD}(r_{\rm{b}}) < t_{\rm break}$, where
\begin{align}
t_{\rm dyn,CBD}(r_{\rm{b}})&=\left(\frac{Gm}{r_b^3}\right)^{-1/2}\nonumber\\
&\sim 8\times 10^4~{\rm{s}}\left(\frac{m}{150~\Msun}\right)^{-1/2}
\left(\frac{r_{\rm{b}}}{5\times 10^{12}~{\rm cm}}\right)^{3/2}\,.
\end{align}
The condition ($t_{\rm dyn,CBD}(r_{\rm{b}}) < t_{\rm break}$) is not satisfied for short $t_{\rm break}$ such as in the case of GW150914-GBM, where we assume $f_{\rm acc}\sim 1$, 
while that is satisfied 
in the case of ZTF19abanrhr with 
$v_{\rm rk}\gtrsim 100~{\rm{km/s}}$ for $m=150~\Msun$.

On the other hand, when $t_{\rm vis,sh}(r_{\rm{b}})<t_{\rm break}$, 
satisfied in $v_{\rm rk}\gtrsim 300~{\rm{km/s}}$ for $m=150~\Msun$, 
all the mass within $r_{\rm{b}}$ accretes, and the accretion rate is enhanced by a factor of $f_{\rm inc} m_{\rm CBD}(<r_{\rm{b}})/m_{\rm CBD}(<r_{\rm acc})$. 
In this case, 
the accretion rate may be even reduced for $r_{\rm acc}\gg r_{\rm{b}}$ 
due to the ejection of gas beyond $r_{\rm{b}}$, 
which is the case when $v_{\rm rk}\gtrsim 2000~{\rm{km/s}}$ for $m=150~\Msun$ (see Eqs.~\ref{eq:rb} and \ref{eq:t_vis}). 
Since its effect is uncertain, 
we assume that the influence on the accretion rate by the recoil shock is effectively taken into account by enhancing or reducing $f_{\rm acc}$.

Here, the recoil velocity for GW190521 is preferred to be $\sim 300~{\rm{km/s}}$ or 
around it ($\sim 100$--$1000~{\rm{km/s}}$) 
 \citep{LIGO20_GW190521_astro}. 
Since $t_{\rm vis}(r_{\rm{b}})\sim 0.1~{\rm yr}$ is comparable to $t_{\rm delay}\sim t_{\rm break}\sim 18~{\rm day}$ inferred for ZTF19abanrhr, we expect that $f_{\rm acc}\sim f_{\rm inc}\gtrsim 1$, which is predicted to be realized unless $v_{\rm rk}\gtrsim 2000~{\rm{km/s}}$. 
Since $f_{\rm acc}$ is an uncertain quantity, 
when we discuss the association of ZTF19abanrhr
we treat it 
as a parameter whose value is adjusted to match the observational data for 
ZTF19abanrhr.

Note that shocked gas, highly pressurized due to the high jet power (high $f_{\rm acc}$), does not significantly eject the circum-black hole disk and reduce the accretion rate onto the black hole, since the dependence of the truncation radius of the circum-black hole disk on the accretion rate is weak (with a power-law index of 8/53, \citealt{Tagawa2022_BHFeedback}), 
where the truncation radius is the radius beyond which the circum-black hole disk is ejected by the cocoon feedback. 
Also, a hollow cavity, predicted to exist around the majority of the black holes 
 \citep{Kimura2021_BubblesBHMs,Tagawa2022_BHFeedback}, 
further weakens this effect.

\section{Telescope Specifications}
\label{sec:telescopes}

The properties of telescopes, which would be useful for detecting the electromagnetic counterparts predicted by our model, are listed in Table~\ref{table:telescope}.

\begin{table*}
\begin{center}
\caption{The name and properties of telescopes appropriate for detecting electromagnetic counterparts }
\label{table:telescope}
\hspace{-5mm}
\begin{tabular}{c|c|c|c}
\hline 
Telescope name & Photon energy
& Sensitivity $[{\rm erg/s/cm^2}]$& Field of view$[{\rm sr}]$\\
\hline\hline
JWST& $\sim 0.04$--$2~[{\rm{eV}}]$&$\sim 10^{-17}$--$10^{-15}$ for $t_{\rm int}\sim10^4~{\rm{s}}$&$\sim 10^{-6}$\\\hline
WISE& $\sim 0.05$--$0.4~[{\rm{eV}}]$&$\sim 10^{-13}$--$10^{-12}$ for $t_{\rm int}\sim10^4~{\rm{s}}$&$\sim 10^{-4}$\\\hline
Roman space telescope \citep{Spergel_2015}
& $\sim 0.6$--$1~[{\rm{eV}}]$&$\sim 4\times 10^{-16}$ for $t_{\rm int}\sim10^4~{\rm{s}}$&$\sim 0.6$\\\hline
ZTF \citep{Bellm_2018}& 
$\sim 1.4$--$3.1~[{\rm{eV}}]$
&$\sim 10^{-13}$ for $t_{\rm int}\sim30~{\rm{s}}$
&0.01\\\hline
Vera Rubin \citep{Ivezic2019}&
$\sim 1.2$--$3.9~[{\rm{eV}}]$
&$\sim 10^{-15}$ for $t_{\rm int}\sim40~{\rm{s}}$&0.003\\\hline
Subaru/HSC \citep{Aihara2018}& 
$\sim 1$--$3~[{\rm{eV}}]$
&$\sim 10^{-16}$--$10^{-15}$ for $t_{\rm int}\sim10^3~{\rm{s}}$
&
0.0005
\\\hline
Tomo-e Gozen \citep{Tomoe2018}& 
$\sim 1.7$--$3.4~[{\rm{eV}}]$
&$\sim 2\times 10^{-13}$ 
for 
$t_{\rm int}\sim 100~{\rm{s}}$
&0.006
\\\hline
Chandra& $\sim 0.2$--$10~[{\rm keV}]$&$\sim10^{-14}$ for $t_{\rm int}\sim 2\times 10^4~{\rm{s}}$&$6\times 10^{-5}$\\\hline
XMM-Newton \citep{Jansen2001}& $\sim 0.4$--$3~[{\rm keV}]$&$\sim10^{-14}$ for $t_{\rm int}\sim 10^4~{\rm{s}}$&$8\times 10^{-5}$\\\hline
HiZ-GUNDAM \citep{Yonetoku2020c}& $\sim 0.4$--$4~[{\rm keV}]$&$\sim10^{-11}$ for $t_{\rm int}\sim10^4~{\rm{s}}$&1.2\\\hline
Einstein Probe \citep{Yuan2015}& $\sim 0.5$--$4~[{\rm keV}]$&$\sim3\times 10^{-11}$ for $t_{\rm int}\sim 10^3~{\rm{s}}$&1.0\\\hline
MAXI \citep{Matsuoka2009_MAXI}& $\sim 2$--$30~[{\rm keV}]$&$\sim7\times 10^{-11}$ for $t_{\rm int}\sim 6\times 10^5~{\rm{s}}$&0.07\\\hline
NuSTAR \citep{Harrison2013}
& $\sim 10$--$30~[{\rm keV}]$&$\sim 10^{-14}$ for $t_{\rm int}\sim 10^6~{\rm{s}}$&$3\times10^{-5}$
\\\hline
FORCE \citep{Mori16_FORCE}
& $\sim 1$--$80~[{\rm keV}]$&$\sim 10^{-14}(t_{\rm int}/10^5~{\rm{s}})^{-1}$&$10^{-5}$\\\hline
{\it Swift} BAT \citep{Barthelmy2005,Tueller2010}& $\sim 15$--$150~[{\rm keV}]$&$\sim 10^{-8}(t_{\rm int}/1~{\rm{s}})^{-1/2}$&1.4\\\hline
{\it Fermi} GBM \citep{Meegan2009}& $\sim 8$--$4000~[{\rm keV}]$&$\sim 10^{-8}$--$10^{-6}$ for $t_{\rm int}\sim1~{\rm{s}}$&$\sim 4\pi$\\\hline
{\it INTEGRAL} SPI-ACS \citep{Winkler2003}& $\sim 75$--$2000~[{\rm keV}]$&$\sim 10^{-7}$--$10^{-6}$ for $t_{\rm int}\sim1~{\rm{s}}$&$\sim 4\pi$\\\hline
\end{tabular}
\end{center}
\end{table*}

\bibliographystyle{aasjournal}

\bibliography{agn_bhm}

\end{document}